\documentclass[aps,amsmath,notitlepage,twocolumn,amssymb,prb,longbibliography]{revtex4-1}

\usepackage[pdftex]{graphicx} 
\usepackage{epstopdf}
\usepackage{verbatim}
\usepackage{color}
\usepackage{subfigure}
\usepackage{tabularx}
\usepackage{amsfonts}
\usepackage{wasysym}
\usepackage{bm}

\usepackage[colorlinks=true,citecolor=red,urlcolor=blue]{hyperref}

\definecolor{darkred}{rgb}{0.847059, 0.141176, 0.164706}
\definecolor{darkgreen}{rgb}{0,0.4,0}
\definecolor{darkblue}{rgb}{0.254902, 0.411765, 0.882353}

\newcolumntype{C}[1]{>{\centering\let\newline\\\arraybackslash\hspace{0pt}}m{#1}}

\begin{document}
\title{Extended Coulomb liquid of paired hardcore boson model on a pyrochlore lattice}
\author{Chun-Jiong Huang$^{1,2,3}$}
\author{Changle Liu$^{4,7}$}
\author{Ziyang Meng$^{5}$}
\author{Yue Yu$^{4,6,7}$}
\author{Youjin Deng$^{1,2,3}$}
\email{yjdeng@ustc.edu.cn}
\author{Gang Chen$^{4,6,7}$}
\email{gangchen.physics@gmail.com}

\affiliation{$^{1}$Shanghai Branch, National Laboratory for Physical Sciences at 
Microscale and Department of Modern Physics, University of Science and 
Technology of China, Shanghai, 201315, China}
\affiliation{$^{2}$CAS Center for Excellence and Synergetic Innovation Center in 
Quantum Information and Quantum Physics, University of Science and Technology of China, Hefei, Anhui 230026, China}
\affiliation{$^{3}$CAS-Alibaba Quantum Computing Laboratory, Shanghai, 201315, China}
\affiliation{$^{4}$State Key Laboratory of Surface Physics and Department of Physics, 
Fudan University, Shanghai, 200433, China}
\affiliation{$^{5}$Beijing National Laboratory for Condensed Matter Physics, Institute of Physics, CAS, Beijing 100190, China}
\affiliation{$^{6}$Center for Field Theory \& Particle Physics, Fudan University, Shanghai, 200433, China}
\affiliation{$^{7}$Collaborative Innovation Center of Advanced Microstructures, Nanjing University, Nanjing, 210093, China}

\date{\today}

\begin{abstract}
There is a growing interest in the $U(1)$ Coulomb liquid in both quantum 
materials in pyrochlore ice and cluster Mott insulators and cold atom systems. 
We explore a paired hardcore boson model on a pyrochlore lattice. 
This model is equivalent to the XYZ spin model that was proposed 
for rare-earth pyrochlores with ``dipole-octupole'' doublets. 
Since this model has no sign problem for quantum Monte Carlo (QMC) 
simulations in a large parameter regime, we carry out both analytical   
and QMC calculations. We find that the $U(1)$ Coulomb liquid is 
quite stable and spans a rather large portion of the phase diagram 
with boson pairing. Moreover, we numerically find thermodynamic 
evidence that the boson pairing could induce a possible $\mathbb{Z}_2$ 
liquid in the vicinity of the phase boundary between Coulomb liquid 
and $\mathbb{Z}_2$ symmetry-broken phase. Besides the materials' 
relevance with quantum spin ice, we point to quantum simulation 
with cold atoms on optical lattices.
\end{abstract}

\maketitle

\noindent The search of exotic quantum phases with quantum number 
fractionalization and emergent gauge structure has been an active 
subject in modern condensed matter physics. One theoretical route 
in the field is to start from the exotic phase itself and construct 
solvable models. These models are often contrived and not quite 
realistic~\cite{KITAEV20032,levinwen,hermele,GuWangWen}. One exception 
is the exactly solvable Kitaev model on the honeycomb lattice~\cite{Kitaevhoneycomb} 
whose physical relevance to the iridate materials was later pointed out by 
G. Jackeli and G. Khaliullin~\cite{PhysRevLett.102.017205}. The opposite route 
is to start from the realistic physical systems and build up relevant models 
from the physical degrees of freedom. Both routes have been quite fruitful. 
The latter route faces several major obstacles. Firstly, constructing 
a relevant physical model itself is not often straight-forward. 
Secondly, these strongly interacting models often cannot be solved 
in a controlled manner. Occasionally, certain realistic models, 
such as the square lattice Heisenberg model
for the cuprates, may be solved but yield a bit mundane and known 
results, and are thus of limited theoretical value for our understanding 
of strongly correlated quantum matters. Therefore, a physically 
relevant model, that can be solved in a controlled manner and at 
the same time gives non-trivial quantum phases, is highly valuable 
in the study of strongly correlated quantum matters.

\begin{figure}[b]
\includegraphics[width=\columnwidth]{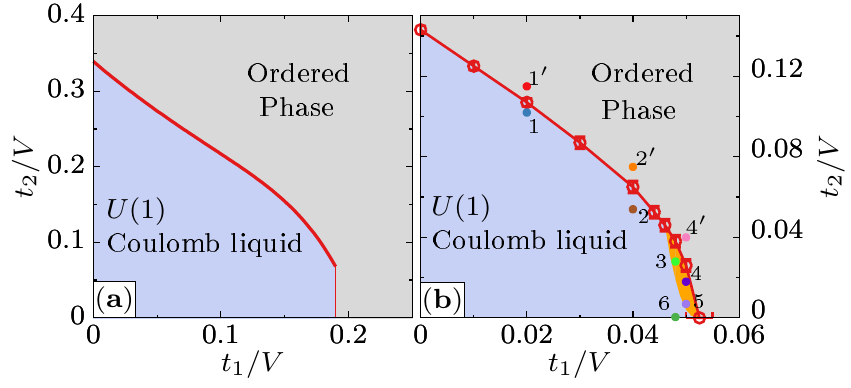}
\caption{{\bf{The phase diagrams of paired hardcore boson model}}. 
(a) The (theoretical) mean-field phase diagram. Thick (thin) line
indicates a first order (continuous) transition.
(b) The QMC phase diagram. The detailed properties 
of the specific data points in the plot are presented in 
Figure~\ref{fig4} and Figure~\ref{fig5}.}
\label{fig1}
\end{figure}

The XYZ spin model, that was derived from the microscopics 
of dipole-octupole doublets on the pyrochlore lattice
and on the triangular lattice by one of us and collaborators 
in Refs.~\onlinecite{Huang2014,Yaodong2016}, is a rare example 
that overcomes the major obstacles of the second route. 
It was suggested that this model on the pyrochlore lattice 
could stabilize a $U(1)$ Coulombic liquid~\cite{PhysRevB.69.064404,Henley} 
and may stabilize a $\mathbb{Z}_2$ spin liquid in parts of its 
phase diagram~\cite{Huang2014}. The $U(1)$ Coulombic liquid 
is an exotic quantum state that is described by compact 
quantum electrodynamics with emergent 
quasiparticles~\cite{PhysRevB.69.064404,Henley} and has 
found relevance in pyrochlore quantum ice materials~\cite{PhysRevLett.98.157204,PhysRevX.1.021002,PhysRevB.90.214430,Sungbin2012,PhysRevLett.108.037202,PhysRevB.86.075154,PhysRevLett.108.247210,Huang2014,PrHfO,PhysRevB.94.205107,PhysRevLett.118.087203,PhysRevLett.118.107206,PhysRevB.96.125145}
and cluster Mott insulators~\cite{PhysRevLett.113.197202,PhysRevB.97.035124,PhysRevB.93.245134,Chen2015,PhysRevB.96.054405}. Besides the non-trivial ground states, 
it was also pointed out~\cite{Huang2014} 
that our model does not have a sign problem for quantum Monte Carlo 
(QMC) simulation in a large parameter regime, and in fact, it is 
the case for any lattice~\cite{Yaodong2016}. An extension of this 
model to the kagom\'{e} lattice by dimensional reduction from the 
pyrochlores with magnetic fields was later pursued
numerically~\cite{Melko2015}. Our model was first proposed for
various Nd-based pyrochlore materials~\cite{Huang2014,PhysRevB.91.174416,
PhysRevB.92.224430,PhysRevB.92.184418,PhysRevB.92.144423,PhysRevLett.115.197202,
PhysRevB.94.104430,PhysRevB.94.064425,PhysRevB.95.224420,PhysRevB.95.134420}, 
and was recently suggested for a Ce-based pyrochlore spin liquid candidate  
Ce$_2$Sn$_2$O$_7$~\cite{PhysRevB.95.041106,PhysRevLett.115.097202}. 
Thus, the XYZ model becomes a rare model that describes real physical 
systems, supports non-trivial quantum phases, and can be solved in a 
controlled manner in a large parameter regime. Inspired by these compelling 
properties of the XYZ model~\cite{Huang2014,Yaodong2016,PhysRevB.95.041106}, 
we carry out both theoretical analysis and numerical calculation 
to establish the phase diagram of this model on the pyrochlore lattice. 
We show that the $U(1)$ Coulomb liquid covers a rather large portion 
of the phase diagram. In addition, the physical boson pairing may render
new fates to the emergent spinon-gauge coupling in the $U(1)$ Coulomb 
liquid~\cite{PhysRevLett.108.037202,Sungbin2012}. We find the 
thermodynamic evidence for the possible existence of a $\mathbb{Z}_2$ 
liquid state out of the $U(1)$ Coulomb liquid via an internal Anderson-Higgs' 
mechanism by the spinon pairing. 
\\

\noindent{\bf Results}\\
\noindent{\bf{\small The paired hardcore boson model.}} We start from the paired 
hardcore boson model on the pyrochlore lattice, where the    
Hamiltonian is given as 
\begin{eqnarray}
H &=& \sum_{\langle ij \rangle}
\big[(-t_1^{} b^{\dagger}_i b^{}_j - t_2^{} b^{\dagger}_i b^{\dagger}_j  + h.c.)
     + V n^{}_i n^{}_j \big]. 
\label{eq1}
\end{eqnarray}
Here, $b^{\dagger}_i$ ($b_i^{}$) creates (annihilates) one boson 
at the lattice site $i$, and $n^{}_i \equiv b_i^{\dagger} b^{}_i$ 
is the boson occupation number. This model differs from the usual 
hardcore boson model~\cite{PhysRevB.69.064404,PhysRevLett.100.047208,
PhysRevLett.105.047201,PhysRevLett.108.067204,Chen2015} 
by having an extra boson pairing term. Previous theoretical works and   
numerical efforts on the hardcore boson model without the boson pairing 
have established the presence of the $U(1)$ Coulomb liquid ground state 
that supports the gapless $U(1)$ gauge photon and fractionalized 
excitations~\cite{PhysRevLett.100.047208,PhysRevLett.115.077202}. 
The main purpose of this work is to understand the role of this 
boson pairing on the phase diagram of the paired hardcore boson model. 

This hardcore model has a strong physical motivation. This model 
is identical to the XYZ spin model via the standard mapping 
${b_i^{} \equiv S^-_i}, {n_i^{} \equiv S^z_i + 1/2}$. 
The spin model was derived as a generic and realistic model 
that describes the interaction between the so-called 
``dipole-octupole doublets'' on the pyrochlore lattice~\cite{Huang2014,Yaodong2016,PhysRevB.95.041106}. 
The boson pairing naturally arises from the spin-orbit 
entanglement of the dipole-octupole doublets. In the end of 
this work, we further mention the relevance with the cold-atom 
systems that have been proposed~\cite{PhysRevLett.114.173002,
PhysRevLett.95.040402}. Due to the boson pairing, the 
global $U(1)$ symmetry is absent and the total boson particle 
number is not conserved, but the Hamiltonian remains invariant 
under a global $\mathbb{Z}_2$ (or Ising) symmetry transformation 
with ${b_i^{} \rightarrow - b_i^{}}, {b_i^{\dagger} \rightarrow 
- b_i^{\dagger}}$. Throughout this work, we work on the regime 
with an average $1/2$-boson filling. In the following, we first 
carry out the theoretical analysis and provide the physical 
understanding of the internal and emergent gauge structure 
and fractionalized excitations of this model, and then implement 
the large-scale QMC simulation to confirm the theoretical expectation. 
\\

\noindent{\bf{\small {The internal gauge structure and phase diagram.}}}
Since the hardcore boson model without pairing is equivalent to
the XXZ spin model and has been extensively studied~\cite{PhysRevB.69.064404,PhysRevLett.100.047208,
PhysRevLett.105.047201,PhysRevLett.108.067204}, we 
briefly explain the ground state in the limit with 
${t_2 = 0}$. When the hopping $t_1$ is greater than 
a critical value, the bosons are simply condensed and 
form a superfluid by breaking the global $U(1)$
symmetry. In the opposite case when $t_1$ is less
than a critical value, the system would form a  
$U(1)$ Coulomb liquid with an emergent $U(1)$ 
gauge structure and fractionalized excitations. 
Note the emergent $U(1)$ gauge structure in the 
$U(1)$ Coulomb liquid has nothing to do with the 
global $U(1)$ symmetry of the model in the XXZ limit. 
Due to the emergent non-locality of the underlying
$U(1)$ gauge structure, the Coulomb liquid in the 
small $t_1$ regime is robust against any small and local 
perturbation such as the weak $t_2$ boson pairing. 

The $U(1)$ Coulomb liquid in the phase diagram  
can also be established from the limit with ${t_1 = 0}$.
As we elaborate in the Supplementary materials, 
a sixth order degenerate perturbation theory in the      
$t_2$ pairing is needed to generate the three-boson 
hopping on the perimeter of the elementary hexagon 
of the pyrochlore lattice. It is this three-boson collective
hopping that allows the system to fluctuate quantum 
mechanically within the extensively degenerate ground 
state manifold (or spin ice~\cite{PhysRevB.69.064404,Bramwell2001,
Castelnovo2008,0034-4885-77-5-056501} manifold in the spin language) 
of the predominant boson interaction and lead to the $U(1)$ 
Coulomb liquid. When both $t_1$ and $t_2$ are present 
and remain small, similar perturbative treatment
again leads to $U(1)$ Coulomb liquid. Therefore, 
we expect the $U(1)$ Coulomb liquid to appear as 
the ground state when both $t_1$ and $t_2$ are 
reasonably smaller than $V$.

\begin{table}[t]
\begin{tabular}{cccc}
\hline\hline
Properties      & $U(1)$ liquid & $\mathbb{Z}_2$ liquid &  Ordered phase 
\\\hline 
gap or not      &  gapless  &  gapped   &  gapped     \\ 
Low-$T$ $C_v$   & power-law & activated &  activated \\   
$\langle \tilde{n}_i \tilde{n}_j \rangle $  & power-law & expo decay & expo decay \\
$\langle b_i^\dagger b_j^{} \rangle $ &  expo decay & expo decay & long-range order \\
\hline\hline
\end{tabular}
\caption{{\bf{The physical properties of different phases.}} 
Here `expo' refers to `exponentially', and ${\tilde{n}_i \equiv {n_i - \frac{1}{2}}}$.
The ordered phase in the upper right region of the phase diagram in Figure~\ref{fig1} 
breaks the global $\mathbb{Z}_2$ (or Ising) symmetry. 
}
\label{tab1}
\end{table}

To establish the phase diagram, we first realize that the 
system favors a ferromagnetic order with $\langle S^x\rangle
\equiv \langle b+b^{\dagger}\rangle /2 \neq 0$
when ${t_1,t_2 \gg V}$ and ${t_1, t_2 > 0}$. 
Moreover, the phases for ${t_2>0}$ and ${t_2 <0}$ are 
related under the transformation $b \rightarrow i b,
b^\dagger \rightarrow -i b^{\dagger}$. To reveal
the connection between the Coulomb liquid and the 
ordered phases, we view the Coulomb liquid as the 
parent phase and implement the spinon-gauge 
construction~\cite{PhysRevLett.108.037202,Sungbin2012} 
for the hardcore boson operators that is appropriate 
for the Coulomb liquid phase,
\begin{eqnarray}
&& b^\dagger_i \equiv \frac{1}{2} \Phi^{\dagger}_{\boldsymbol r} 
\Phi^{}_{{\boldsymbol r}'} 
e^{i A_{{\boldsymbol r}{\boldsymbol r}'} } , \quad
 \sum_{i \in \text{tet}_{\boldsymbol r}} n_i^{} 
= \eta_{\boldsymbol r}^{} Q_{\boldsymbol r}^{} + 2 ,
\end{eqnarray}
where $\Phi^{\dagger}_{\boldsymbol r}$ 
($\Phi^{}_{\boldsymbol r}$) 
creates (annihilates) a spinon at the center 
(labeled by `${\boldsymbol r}$') of the tetrahedron (`tet$_{\boldsymbol r}$'),
and $\eta_{\boldsymbol r} =\pm1$ for two sublattices of the diamond
lattice formed by the tetrahedral centers. 
As we explain in details in Methods, the paired hardcore boson model becomes
\begin{eqnarray}
&&H = \sum_{\boldsymbol r} \frac{V}{2} Q^2_{\boldsymbol r} - \frac{t_1^{} }{4}
\sum_{\langle {\boldsymbol r}{\boldsymbol r}' \rangle} 
\Phi^\dagger_{\boldsymbol r} \Phi^{}_{{\boldsymbol r}'} 
e^{- i (A_{ {\boldsymbol r} {\boldsymbol r}''  } + A_{{\boldsymbol r}'' {\boldsymbol r}'}) }
\nonumber \\
&& 
- \frac{t_2^{}}{8} \sum_{\langle {\boldsymbol r} {\boldsymbol r}' \rangle}
\sum_{\langle {\boldsymbol r} {\boldsymbol r}'' \rangle}
\big[\Phi^\dagger_{\boldsymbol r} \Phi^\dagger_{\boldsymbol r} 
\Phi^{}_{{\boldsymbol r}'}\Phi^{}_{{\boldsymbol r}''} 
e^{-i ( A_{ {\boldsymbol r} {\boldsymbol r}' } + A_{ {\boldsymbol r} {\boldsymbol r}'' }  )}
+ h.c. \big].
\end{eqnarray}
It is noticed that the boson pairing mediates the
spinon interaction in the spinon-gauge formulation~\cite{Sungbin2012}. 
The spinon interaction may induce pairing between these fractionalized 
degrees of freedom and thus gap out the continuous part of the 
internal $U(1)$ gauge field via an internal Anderson-Higgs' mechanism. 
Through the standard mean-field analysis, we do not actually find any
pairing instability within the Coulomb liquid phase in the mean-field phase 
diagram (see Figure~\ref{fig1}). This can be a mean-field artifact. 
Nevertheless, the mean-field theory does give a large region for 
$U(1)$ Coulomb liquid in the phase diagram. 
\\

\noindent{\bf{\small {QMC algorithm}.}} To examine the theoretical understanding, 
we perform the worm-type QMC algorithm~\cite{Prokofev1998,Prokofev1998b} 
to simulate the model in Eq.~\eqref{eq1}. Since there are both boson 
hopping and pairing terms in the model, the typical worm QMC is no longer 
sufficient and a new update scheme is needed, which we outline here 
and more details can be found in Methods.

\begin{figure}[t]
\includegraphics[width=7.5cm]{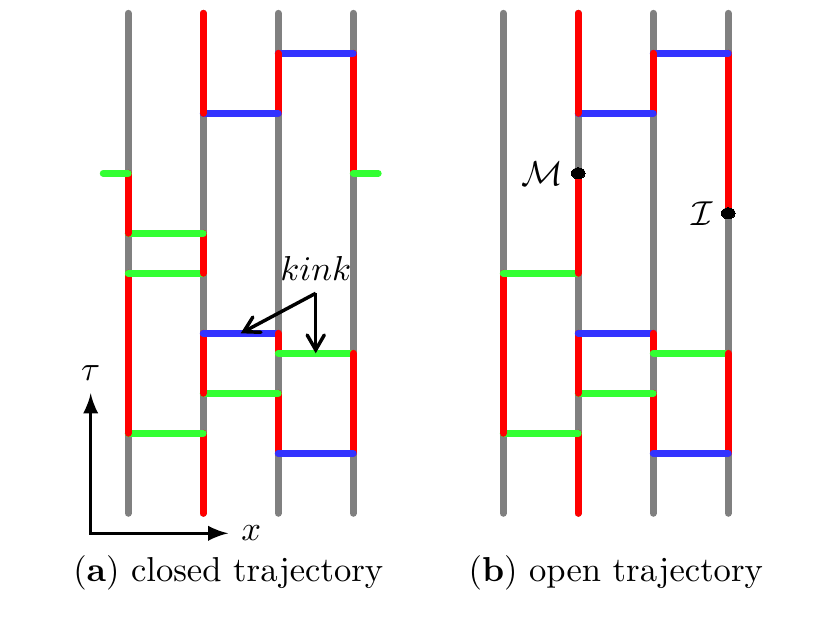}
\caption{{\bf{The worldline trajectories 
under imarginary time evolution.}}
The red (gray)
vertical segments with different colors along the worldlines 
refer to bosons occupied (vacuum) states, and the boson hopping 
and pairing are represented by the green and blue horizontal 
links, respectively.  
(a) is the closed trajectory that contributes to the partition 
	function $\mathcal{Z}$ and (b) is the open trajectory with two worldline 
	discontinuities, 
	$\mathcal{I}$ and $\mathcal{M}$, that belong to the 
	configuration space of 
	$\mathcal{G} (i,\tau_{_\mathcal{I}};j,\tau_{_\mathcal{M}}) $. 
 }
\label{fig2}
\end{figure}

We first express the partition function via Trotter 
product expansion in the imaginary time. We split the Hamiltonian 
into the free part ${ K=\sum_{\langle ij\rangle}[-t_1 b_i^\dagger b_j 
-t_2 b_i^\dagger b_j^\dagger+h.c.]  }$ and the interaction part 
$U={\sum_{\langle ij\rangle}V n_in_j-\mu\sum_i n_i}$, and expand the 
partition function with respect to the interaction part (using 
occupation basis that is denoted as ${|\alpha\rangle}$), where the 
grand canonical ensemble for the bosons is used by introducing the 
chemical potential to the bosons. The partition function is given as
\begin{eqnarray}
\mathcal{Z} &= & \text{Tr}\left[e^{-\beta \mathcal{H}}\right] %
= \sum\limits_{\{\alpha_0\}}\langle\alpha_0|e^{-\beta
	\mathcal{H}}|\alpha_0\rangle
	\nonumber \\
&= & 
\lim_{\scriptstyle{d\tau=\frac{\beta}{n}}\atop\scriptstyle{n\rightarrow 
\infty}}\sum\limits_{\scriptstyle{ \{\alpha\} }\atop\scriptstyle{\alpha_n=\alpha_0}} %
\langle\alpha_n|e^{-\mathcal{H}d\tau}|\alpha_{n-1}\rangle %
\cdots \langle\alpha_1|e^{-\mathcal{H}d\tau}|\alpha_0\rangle
\nonumber \\
&= & \sum\limits_{\{\alpha\}}\sum_{k_1,k_2=0}^{\infty} %
\int_0^\beta \cdot \cdot \int_{\tau_{_{k-1}}}^\beta %
\prod_{i=1}^{k} d\tau_i t_1^{k_1}t_2^{k_2}
 e^{{-\int_0^\beta U(\tau)d\tau}},
\label{parfun}
\end{eqnarray}
where ${k\equiv k_1 + k_2}$. Under this representation, the configuration space of the 
partition function~$\mathcal{Z}$ consists of all trajectories with 
closed worldlines (see Figure~\ref{fig2}(a)), where ``closed" 
refers to a periodic boundary condition with ${|\alpha(0)\rangle=|\alpha(\beta)\rangle}$. 
Due to the off-diagonal operator $K$, a boson can hop from one site to 
its neighbors via $-t_1 b^\dagger_i b^{}_j$, or one pair of 
bosons can be created or annihilated at the same imaginary time 
through $-t_2 b^\dagger_i b^\dagger_j$ or $-t_2 b_i b_j$, and these 
two processes are dubbed hopping and pairing kinks, 
respectively. The numbers of such kinks are given by $k_1$ and $k_2$ 
in Eq.~\eqref{parfun}. To evaluate the dynamical properties, 
we further define a particular Green's function as
\begin{eqnarray}
\label{greenfun}
&& \mathcal{G}(i,\tau_{_\mathcal{I}};j,\tau_{_\mathcal{M}}) 
\nonumber \\
&&
\equiv 
\textrm{\large Tr}\, \textrm{T}_\tau 
\Big[\big[ b_i^\dagger(\tau_{_\mathcal{I}})  +  b_i^{}(\tau_{_\mathcal{I}})  \big]
 \big[ b_j^{\dagger}(\tau_{_\mathcal{M}})+  b_j^{}(\tau_{_\mathcal{M}})  \big]
e^{-\beta
\mathcal{H}} \Big]  .
\end{eqnarray}
As we show in Figure~\ref{fig2}(b), $\mathcal{G}(i,\tau_{_\mathcal{I}};j,\tau_{_\mathcal{M}})$
introduces open trajectories that contain 
two worldline discontinuities ``$\mathcal{I}$" and ``$\mathcal{M}$",
and $(i,\tau_{_\mathcal{I}})$ and $(j,\tau_{_\mathcal{M}})$ are the spatial, 
temporal locations of two worldline discontinuities. Shifting the discontinuities 
in space and time produces a series of trajectories. This is a crucial benefit 
of the \textit{worm-type} algorithm that we can calculate the Green's function as 
efficiently as other thermodynamic quantities.

All closed and open trajectories constitute the total configuration space of 
the \textit{worm-type} algorithm. Through three types of update procedures 
with certain probabilities, we can produce a Markov chain of different 
trajectories that walk in the total configuration space randomly. These 
procedures are classified as: (1) creation and annihilation of two worldline 
discontinuities, $\mathcal{I}$ and $\mathcal{M}$, (2) shift of $\mathcal{I}$ in 
time and (3) creation and deletion of kinks. 
The procedures of creating (deleting) kinks can be further divided into four specific 
ones with certain ratios. The thermodynamic properties are measured 
in the closed space, and histograms of the Green's function are counted in the open space.
\\

\begin{figure}[t]	
\includegraphics[width=\columnwidth]{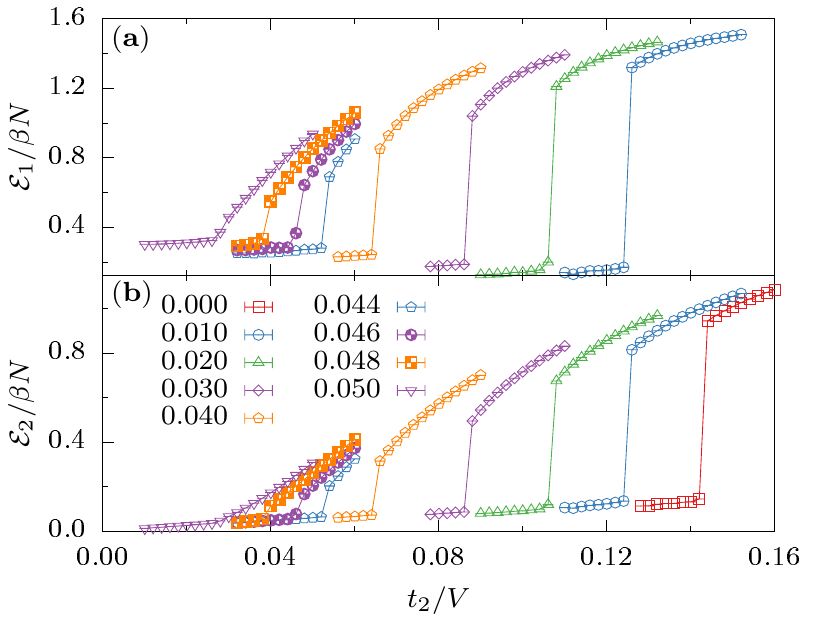}
\caption{{\bf{The hopping and pairing kinks by varying couplings.}}
The legend suggests the values of $t_1/V$. The system size in these 
simulations is ${N=4\times 8^3}$. When ${t_1 \leq 0.048 V}$ the 
curves of both kink types are clearly discontinuous, indicating 
strongly first order transitions. For ${t_1=0.048V}$ and ${t_1=0.05V}$, 
a weakly first order phase transition is more likely.}
\label{fig:twokinks}
\end{figure}

\noindent{\bf{\small QMC results for different phases.}} 
To comply with the XYZ model, we further set the chemical potential 
${\mu \equiv 3V}$ in the following calculations. To determine the phase 
boundary between the disordered liquid phases and the ordered phase, 
we monitor the first order derivative of the free energy 
over the parameters $t_1$ and $t_2$ with 
\begin{eqnarray}
\mathcal{E}_{1} = {\partial\, {\ln {\mathcal Z} }}/{\partial t_1} &=& 
{\langle k_1 \rangle}/{t_1}, \\
\mathcal{E}_{2} = {\partial\, {\ln {\mathcal Z} }}/{\partial t_2} &=& 
{\langle k_2 \rangle}/{t_2}.
\label{kinkdef}
\end{eqnarray}
We simulate these values by varying $t_2$ for fixed $t_1$'s with the 
system size ${N=4\times 8^3,\beta=(k_{\text B} T)^{-1}=800}$, where we set 
${V = 1}$ as the energy unit. The numerical phase diagram is presented in 
Figure~\ref{fig1}(b). The transitions are strongly first order at small $t_1$'s 
and are consistent with the theoretical results in Figure~\ref{fig1}(a). 
Moreover, as the system approaches the phase boundary near the horizontal 
axis, the transition becomes weakly first order like. 
In general, the phase boundary in Figure~\ref{fig1}(b) is 
qualitatively consistent with the theoretical one.

\begin{figure}[t]	
\includegraphics[width=\columnwidth]{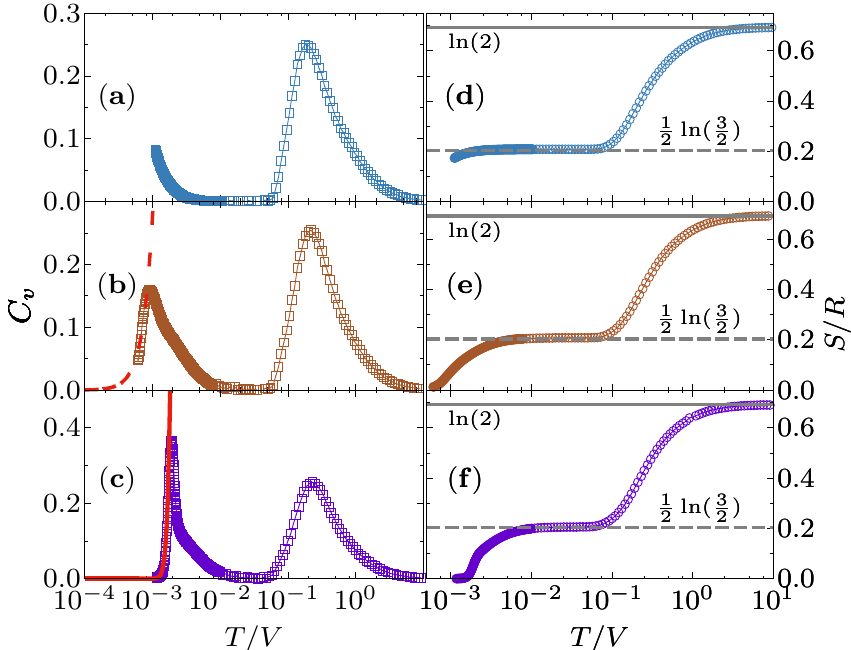}
\caption{{\bf{Heat capacity and entropy density for different couplings.}}
(a), (b) and (c) are heat capacities for the parameter points 1, 2 and 4, 
respectively. (d), (e) and (f) are entropy densities for the parameter 
points 1, 2 and 4, respectively. Solid (dashed) curve is a fit of $T^3$ 
(exponential decaying) behavior. 
}
\label{fig4}
\end{figure}

To understand different phases,  
we probe the thermodynamic properties by measuring the specific 
heat and the entropy for the representative points in Figure~\ref{fig1}(b). 
The results are depicted in Figure~\ref{fig4}.
For the $U(1)$ Coulomb liquid in the pyrochlore ice context~\cite{PhysRevB.69.064404,0034-4885-77-5-056501,Henley}, 
it is well-known that there exist double peaks in the heat 
capacity. The high temperature peak signals the entering 
into the spin ice manifold, while the low temperature 
peak arises from the quantum fluctuation that breaks 
the classical degeneracy of the spin ice manifold. 
Between the two peaks, there is an entropy plateau at the value
of the Pauling entropy
since the system is fluctuating within the ice manifold. 
Below the low temperature peak, the specific heat behaves 
as ${C_v \propto T^3}$ in the zero temperature limit due to the 
gapless $U(1)$ gauge photon~\cite{PhysRevLett.108.037202}. 
For the representative points 1,2 
in Figure~\ref{fig1}(b), the behavior of the specific heat is consistent 
with the $U(1)$ Coulomb liquid (see Figure~\ref{fig4}). 
This gapless excitation is the 
key signature of the emergent gauge dynamics,
and is not related to any continuous symmetry breaking, especially
since there is no symmetry breaking in the disordered regime and the
(generic) model~\cite{Huang2014} does not even have a continuous symmetry.

For the $\mathbb{Z}_2$ liquid, all the excitations are fully gapped.   
Since the spinon pairing is expected to occur at very low energy 
scale, the double peaks in the heat capacity should persist except
that we have an activated behavior of the heat capacity below the
low temperature peak instead of the $T^3$ behavior for the
$U(1)$ Coulomb liquid. Inside the disordered regime of Figure~\ref{fig1}(b),     
we find that the behaviors of ``points 3,4,5''
are consistent with a $\mathbb{Z}_2$ 
liquid (see Figure~\ref{fig4} and Supplementary material). 
This result provides a thermodynamic evidence for 
the presence of a $\mathbb{Z}_2$ liquid phase in the (orange) 
region of the disordered regime. More specifically, the 
thermodynamic gap, that is extracted from the heat capacity 
for the parameter point 4, is ${\sim 0.018V}$. This is
of the same order as the $t_2$ value, suggesting the possible physical 
origin of the $\mathbb{Z}_2$ liquid state. 
As it was noted, 
the $t_2$ term renders an effective interaction between the 
(fractionalized) spinon quasiparticles. When one pair of 
spinons is condensed and individual spinon remains uncondensed, 
the $U(1)$ Coulomb liquid would give way to the $\mathbb{Z}_2$
liquid in a way similar to the superconducting pairing transition 
in a BCS superconductor. More physically, as $t_1/V$ increases inside
the $U(1)$ Coulomb liquid, the spinon gap monotonically decreases, 
and the interaction $t_2$ could lower the spinon pairing energy and overcome the 
reduced two-spinon gap, leading to the $\mathbb{Z}_2$ liquid state. 
In the Supplementary material, we provide more discussion about
the detailed features of the specific heat in Figure~\ref{fig4} 
and discuss the possibility of charge density wave as an alternative 
explanation.


As listed in Table~\ref{tab1}, another important distinction between different 
quantum phases lies in the spatial dependence of correlation functions.   
Here we numerically measure the density-density and the 
boson-boson correlators that are defined as 
\begin{eqnarray}
C_n ({\boldsymbol r}) & \equiv & \langle (n_i-{1}/{2} )%
                      (n_j-{1}/{2} )  \rangle , \\
C_b ({\boldsymbol r}) & \equiv & \langle b^{\dagger}_i b^{}_j \rangle ,
\end{eqnarray}
where ${\boldsymbol r}$ is the spatial separation between the lattice sites $i$ and $j$. 
In the spin language, $C_n$ would correspond to the $S^{z}$-$S^z$ correlator,
while $C_b$ corresponds to the $S^+$-$S^-$ correlator.

\begin{figure}[t]
\includegraphics[width=\columnwidth]{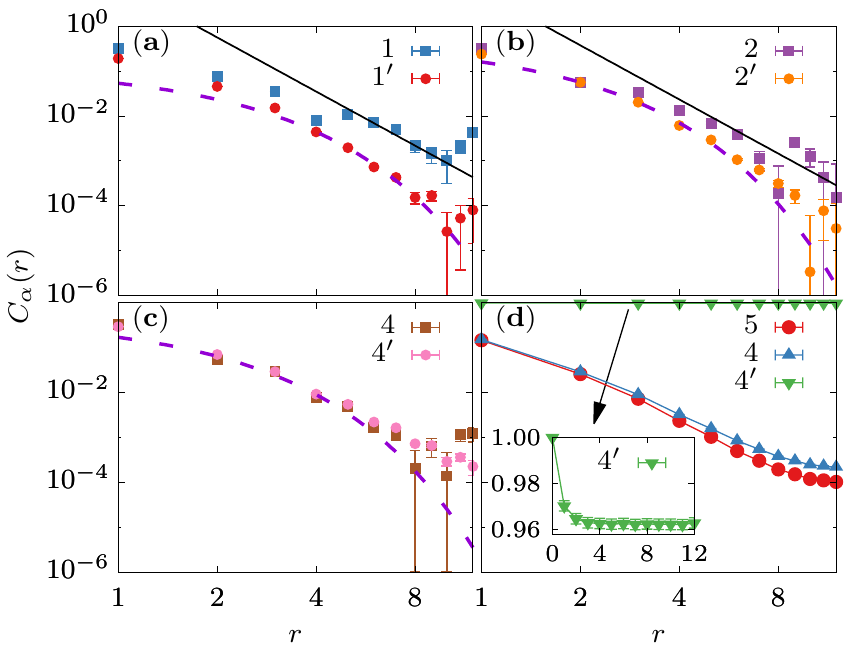}
\caption{{\bf{The density and boson correlators for different couplings.}}
(a), (b) and (c) are (equal-time) density correlators $C_{n}$ for the given parameter 
points in the figures, and (d) lists the boson-boson correlators $C_b$. 
For the $C_{n}$, we have taken the absolute value. The solid (dashed) line 
refers to the behavior of a $1/r^4$ power-law (exponential) decaying.  
}
\label{fig5}
\end{figure}

We first compare the correlations of the $U(1)$ Coulomb liquid 
and those of the ordered state. As we depict in Figure~\ref{fig5}, 
the boson density correlators for the parameter points 1,2
decay as a $1/r^4$ power-law with the distance, and the boson-boson correlators
decay exponentially. This is consistent with the prediction      
from the $U(1)$ Coulomb liquid in which the density correlator 
at long distances and low energies~\cite{PhysRevLett.108.037202,PhysRevB.86.075154} 
is mapped to the $U(1)$ gauge photon modes~\footnote{It 
was recently realized in Ref.~\onlinecite{PhysRevB.96.195127} that at higher energies, 
the density correlator would include the magnetic monopole contribution.} 
and the boson-boson correlator reflects the gapped fractionalized
(spinon) quasiparticles~\cite{PhysRevB.96.085136,PhysRevB.96.195127}. In contrast, for the 
representative parameter point $4'$ inside the ordered state, the boson 
density correlator decays exponentially, and the boson-boson correlator
saturates to a constant since the system develops the  
order in $\langle b \rangle$ by breaking the global $\mathbb{Z}_2$ 
symmetry and simultaneously gives rise to a gap for the density
correlator.

For the $\mathbb{Z}_2$ liquid state, all the correlators should 
decay exponentially with the spatial separations. In our calculation,
we find the boson-boson correlation does indeed decay exponentially.
For the density correlators, despite the thermodynamic gap, we were 
unable to show more convincingly the exponetially decaying behavior 
due to the finite system size in our simulation and the tiny energy gap. 
To resolve this, one may need even larger system sizes to carry out
the simulation in the future work.  
\\

\noindent{\emph{\bf Discussion}}\\
\noindent
We discuss the physical realization of our spin or hardcore boson model.
The solid-state realization has been proposed for the dipole-octupole 
doublets and studied in the previous works by one of us and 
collaborators~\cite{Huang2014,Yaodong2016,PhysRevB.95.041106}. 
Several Nd-based~\cite{PhysRevB.91.174416,
PhysRevB.92.224430,PhysRevB.92.184418,PhysRevB.92.144423,PhysRevLett.115.197202,
PhysRevB.94.104430,PhysRevB.94.064425,PhysRevB.95.224420,PhysRevB.95.134420} 
and Sm-based~\cite{SmTiO} pyrochlore magnets~\cite{RevModPhys.82.53} 
have been proposed to realize the dipole-octupole doublets, though 
most of them seem to support magnetic orders with mixed dipolar and 
octupolar components~\cite{PhysRevB.91.174416,
PhysRevB.92.224430,PhysRevB.92.184418,PhysRevB.92.144423,PhysRevLett.115.197202,
PhysRevB.94.104430,PhysRevB.94.064425,PhysRevB.95.224420,PhysRevB.95.134420}. 
The known example of spin liquid candidate is 
the Ce-based pyrochlore Ce$_2$Sn$_2$O$_7$ where the Ce$^{3+}$ ion 
gives a dipole-octupole doublet~\cite{PhysRevB.95.041106,PhysRevLett.115.097202}. 
Therefore, Ce$_2$Sn$_2$O$_7$ should be a good candidate to examine 
the spin liquid physics of the XYZ spin model.

Beyond the solid state context, the cold atoms on optical lattices 
can be used to realize exotic models such as the paired hardcore 
boson model in this work. In a previous proposal, Ref.~\onlinecite{PhysRevLett.95.040402}
has designed a ring exchange interaction for the bosonic gases via 
a Raman transition to ``molecular'' states on optical lattices 
to simulate the $U(1)$ lattice gauge fields, where this Raman 
coupling has the form $\phi^\dagger b_i b_j$ and $\phi$ refers 
to the ``molecular'' state. More recently, the cold alkali atoms stored 
in optical lattices or magnetic trap arrays were proposed to realize a 
broad class of spin-1/2 models including the XYZ model by admixing 
van der Waals interaction between fine-structure split Rydberg states with 
laser light. Following these early proposals, we suggest two cold-atom 
setups to realize our paired hardcore boson models. In the first setup, 
we closely follow Ref.~\onlinecite{PhysRevLett.95.040402} 
and also propose a resonant coupling of the bosons via 
a Raman transition to a ``molecular'' two-particle state. 
Instead of choosing the original $d$-wave symmetry to simulate the ring
exchange in Ref.~\onlinecite{PhysRevLett.95.040402}, we propose a $s$-wave symmetry
and condense the molecular states $\phi$. Such a design naturally 
gives rise to an (uniform) hardcore boson pairing term 
$\langle \phi^{\dagger} \rangle b_i b_j$ for a given lattice.
For the second setup, one can directly make use of the known
results and methods in Ref.~\onlinecite{PhysRevLett.114.173002} 
and extend to other lattices.

In summary, we have studied a paired hardcore boson model (or 
XYZ spin model in the spin language) on a pyrochlore lattice 
and found the broad existence of extoic quantum ground states.
We make various suggestions for the experimental realizations
in the solid-state and cold-atom contexts. 
\\

\noindent{\emph{\bf Methods}}\\
{\bf\small Mean-field scheme.} We describe the mean-field description 
in some details so that the underlying gauge structure and spinon-gauge    
interaction can be manifest. In the following we use the hardcore boson and
the spin languages interchangeably. 
The degenerate classical spin ice configuration is equivalent to the 
the occupuation configuration of two bosons on each tetrahedron. 
We start with the physical meaning of the boson operators 
$b_i^{}$ and $b_i^\dagger$. From the perturbative analysis 
that is explained in the Supplementary materials for completeness, 
we learn that, the three-boson collective hopping becomes the ``magnetic term'' 
in the $U(1)$ gauge theory Hamiltonian and formulation. Thus, $b_i^{}$ and 
$b_i^\dagger$ would correspond to the vector $U(1)$ gauge link of the $U(1)$ 
quantum electrodynamics from this perspective. Physically, the perturbative 
calculation restricts us to the low-energy classical spin ice manifold and 
`throws' away the high-energy excited states. Clearly, perturbative effective 
Hamiltonian doe not have information about the (spinon) matter field that 
carres $U(1)$ gauge charges. Applying $b^\dagger_i$ breaks the ice rule on 
two neighboring tetrahedra that share the site $i$. Spinon excitations are 
created on the diamond lattice that is formed by the tetrahedral centers. 
Thus, $b_i^{}$ and $b_i^\dagger$ carry two pieces of physical content, and 
the spinon-gauge contruction~\cite{PhysRevLett.108.037202,Sungbin2012}
clearly reflects this.

\begin{figure}[t!]
		\includegraphics[width=0.8\columnwidth]{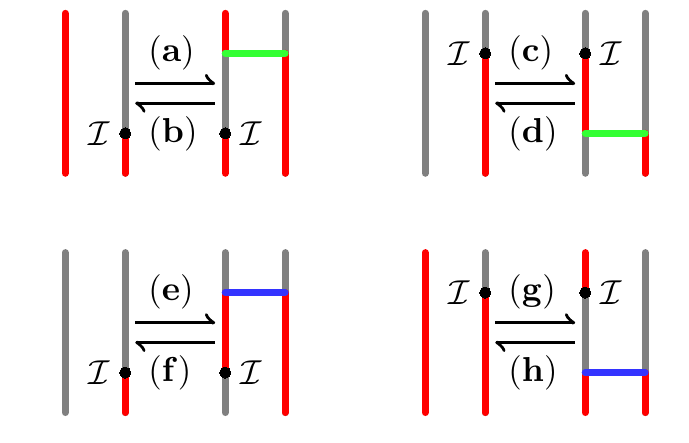}
		\caption{{\bf The procedures of creation and deletion of kink.} The 
			meanings of segments with different colors are identical to the ones in 	
			Figure~\ref{fig1}.}
		\label{pairing}
\end{figure}

In the spinon-gauge formulation in the main text
(that was originally introduced in Refs.~\onlinecite{PhysRevLett.108.037202,Sungbin2012}), 
we have enlarged the physical Hilbert space. To return to 
the physical Hilbert space, a constraint was imposed in the 
main text~\cite{PhysRevLett.108.037202,Sungbin2012}). Since the 
$Q_{\boldsymbol r}$ counts the spinon number, we further have
$[\Phi_{\boldsymbol r}, Q_{\boldsymbol r}] = \Phi_{\boldsymbol r}$
and  
$[\Phi^{\dagger}_{\boldsymbol r}, Q_{\boldsymbol r}^{}] 
= - \Phi^{\dagger}_{\boldsymbol r}$. The spinon-gauge formulation 
of the microscopic Hamiltonian has been introduced in the main text, and 
is written here with more detailed position indices for more readability,
\begin{eqnarray}
H & \simeq & \sum_{\boldsymbol r} \frac{V}{2} Q^2_{\boldsymbol r}  
 - \frac{t_1}{4} \sum_{\boldsymbol r} \sum_{\mu\neq \nu} 
 \Phi^\dagger_{{\boldsymbol r} + \eta_{\boldsymbol r} e_{\mu}} 
  \Phi^{}_{{\boldsymbol r} + \eta_{\boldsymbol r} e_{\nu}} 
  \nonumber \\
 && - \frac{t_2}{8} \sum_{\boldsymbol r} \sum_{\mu \neq \nu} 
  (  \Phi^{\dagger}_{\boldsymbol r} \Phi^{\dagger}_{\boldsymbol r}
 \Phi^{ }_{{\boldsymbol r} + \eta_{\boldsymbol r} e_{\mu} }  
 \Phi^{\dagger}_{{\boldsymbol r} + \eta_{\boldsymbol r} e_{\nu} }
+ h.c.
  ) ,
\end{eqnarray} 
where we set the gauge link ${A_{{\boldsymbol r}{\boldsymbol r}'} = 0}$
since we are dealing with ${t_1>0}$ and zero-flux sector for the spinons
(see Supplementary materials),
$e_{\mu}({\mu=1,2,3,4})$ refers to one of the four nearest-neighbor
vectors on the diamond lattice, and there is a double counting of $\mu$ 
and $\nu$. The mean-field results are obtained by 
systematically decoupling the spinon interaction into
two-spinon terms with self-consistent mean-field conditions.
These procedures are standard and follow closely with
Ref.~\onlinecite{Sungbin2012} that deals with a different model
for non-Kramers doublets on the pyrochlore lattice.
\\

\noindent{\bf\small {Quantum Monte Carlo}.}
We here give some details about our {\it worm-type} algorithm. Procedure 
(1) and	(2) are same as the conventional one. Procedure (3) of creating
(deleting) kinks are consisting of four specific ones: (a)
and (b) are the creation and deletion of hopping kink after $\mathcal{I}$, (c)
and (d) are the creation and deletion of hopping kink before $\mathcal{I}$, 
(e) and (f) are 	the creation and deletion of pairing kink after $\mathcal{I}$ 
and (g) and (h) are the creation and deletion of pairing kink before $\mathcal{I}$.
Procedures (e-g) are introduced due to the pairing term $(b^\dagger b^\dagger+h.c)$ 
so there is no such procedure in conventional Bose-Hubbard model. 
Figure~\ref{pairing} gives the schematic diagram of these procedures.

There is a note worth discussion here. In every {\it worm-type} algorithm 
there is an arbitrary value of $\omega_G$ which defines the relative weight 
of closed and open space. The form of detailed balance equation for Procedure 
(1) is as follows:
\begin{equation}
P_{acc}^{closed\rightarrow open} \omega_G \frac{1}{N_s} W_\mu = 
P_{acc}^{open\rightarrow closed} W_\nu
\end{equation}
where $N_s$ is the number of segments and it is proportional to $N\beta$ 
approximately. $N$ is the number of lattice site and $\beta$ is the 
reciprocal of temperature. We can see that if there is no $\omega_G$ and 
$N\beta$ is very large then the acceptance of changing a open trajectory 
to an closed one will be very small and the algorithm will be very inefficient. 
Here we choose ${\omega_G=N\beta}$ according to the common choice which 
makes ${\omega_G/N_s\sim \text{const.}}$. More details can be found in 
Ref.~\onlinecite{Prokofev1998}.
	\\



\noindent{\bf\small Data availability}\\
{ \noindent The data that support the findings of this study
are available from the corresponding authors (G.C. and Y.J.D.) 
upon request.} 

\vspace{1cm}

\noindent{\emph{\bf References}}
\bibliography{REFS}

\begin{thebibliography}{55}%
\makeatletter
\providecommand \@ifxundefined [1]{%
 \@ifx{#1\undefined}
}%
\providecommand \@ifnum [1]{%
 \ifnum #1\expandafter \@firstoftwo
 \else \expandafter \@secondoftwo
 \fi
}%
\providecommand \@ifx [1]{%
 \ifx #1\expandafter \@firstoftwo
 \else \expandafter \@secondoftwo
 \fi
}%
\providecommand \natexlab [1]{#1}%
\providecommand \enquote  [1]{``#1''}%
\providecommand \bibnamefont  [1]{#1}%
\providecommand \bibfnamefont [1]{#1}%
\providecommand \citenamefont [1]{#1}%
\providecommand \href@noop [0]{\@secondoftwo}%
\providecommand \href [0]{\begingroup \@sanitize@url \@href}%
\providecommand \@href[1]{\@@startlink{#1}\@@href}%
\providecommand \@@href[1]{\endgroup#1\@@endlink}%
\providecommand \@sanitize@url [0]{\catcode `\\12\catcode `\$12\catcode
  `\&12\catcode `\#12\catcode `\^12\catcode `\_12\catcode `\%12\relax}%
\providecommand \@@startlink[1]{}%
\providecommand \@@endlink[0]{}%
\providecommand \url  [0]{\begingroup\@sanitize@url \@url }%
\providecommand \@url [1]{\endgroup\@href {#1}{\urlprefix }}%
\providecommand \urlprefix  [0]{URL }%
\providecommand \Eprint [0]{\href }%
\providecommand \doibase [0]{http://dx.doi.org/}%
\providecommand \selectlanguage [0]{\@gobble}%
\providecommand \bibinfo  [0]{\@secondoftwo}%
\providecommand \bibfield  [0]{\@secondoftwo}%
\providecommand \translation [1]{[#1]}%
\providecommand \BibitemOpen [0]{}%
\providecommand \bibitemStop [0]{}%
\providecommand \bibitemNoStop [0]{.\EOS\space}%
\providecommand \EOS [0]{\spacefactor3000\relax}%
\providecommand \BibitemShut  [1]{\csname bibitem#1\endcsname}%
\let\auto@bib@innerbib\@empty
\bibitem [{\citenamefont {Kitaev}(2003)}]{KITAEV20032}%
  \BibitemOpen
  \bibfield  {author} {\bibinfo {author} {\bibfnamefont {A.Yu.}\ \bibnamefont
  {Kitaev}},\ }\bibfield  {title} {\enquote {\bibinfo {title} {Fault-tolerant
  quantum computation by anyons},}\ }\href {\doibase
  https://doi.org/10.1016/S0003-4916(02)00018-0} {\bibfield  {journal}
  {\bibinfo  {journal} {Annals of Physics}\ }\textbf {\bibinfo {volume}
  {303}},\ \bibinfo {pages} {2 -- 30} (\bibinfo {year} {2003})}\BibitemShut
  {NoStop}%
\bibitem [{\citenamefont {Levin}\ and\ \citenamefont {Wen}(2005)}]{levinwen}%
  \BibitemOpen
  \bibfield  {author} {\bibinfo {author} {\bibfnamefont {Michael~A.}\
  \bibnamefont {Levin}}\ and\ \bibinfo {author} {\bibfnamefont {Xiao-Gang}\
  \bibnamefont {Wen}},\ }\bibfield  {title} {\enquote {\bibinfo {title}
  {String-net condensation: A physical mechanism for topological phases},}\
  }\href {\doibase 10.1103/PhysRevB.71.045110} {\bibfield  {journal} {\bibinfo
  {journal} {Phys. Rev. B}\ }\textbf {\bibinfo {volume} {71}},\ \bibinfo
  {pages} {045110} (\bibinfo {year} {2005})}\BibitemShut {NoStop}%
\bibitem [{\citenamefont {Song}\ and\ \citenamefont {Hermele}(2015)}]{hermele}%
  \BibitemOpen
  \bibfield  {author} {\bibinfo {author} {\bibfnamefont {Hao}\ \bibnamefont
  {Song}}\ and\ \bibinfo {author} {\bibfnamefont {Michael}\ \bibnamefont
  {Hermele}},\ }\bibfield  {title} {\enquote {\bibinfo {title} {{Space-group
  symmetry fractionalization in a family of exactly solvable models with
  ${\mathbb{Z}}_{2}$ topological order}},}\ }\href {\doibase
  10.1103/PhysRevB.91.014405} {\bibfield  {journal} {\bibinfo  {journal} {Phys.
  Rev. B}\ }\textbf {\bibinfo {volume} {91}},\ \bibinfo {pages} {014405}
  (\bibinfo {year} {2015})}\BibitemShut {NoStop}%
\bibitem [{\citenamefont {Gu}\ \emph {et~al.}(2014)\citenamefont {Gu},
  \citenamefont {Wang},\ and\ \citenamefont {Wen}}]{GuWangWen}%
  \BibitemOpen
  \bibfield  {author} {\bibinfo {author} {\bibfnamefont {Zheng-Cheng}\
  \bibnamefont {Gu}}, \bibinfo {author} {\bibfnamefont {Zhenghan}\ \bibnamefont
  {Wang}}, \ and\ \bibinfo {author} {\bibfnamefont {Xiao-Gang}\ \bibnamefont
  {Wen}},\ }\bibfield  {title} {\enquote {\bibinfo {title} {Lattice model for
  fermionic toric code},}\ }\href {\doibase 10.1103/PhysRevB.90.085140}
  {\bibfield  {journal} {\bibinfo  {journal} {Phys. Rev. B}\ }\textbf {\bibinfo
  {volume} {90}},\ \bibinfo {pages} {085140} (\bibinfo {year}
  {2014})}\BibitemShut {NoStop}%
\bibitem [{\citenamefont {Kitaev}(2006)}]{Kitaevhoneycomb}%
  \BibitemOpen
  \bibfield  {author} {\bibinfo {author} {\bibfnamefont {Alexei}\ \bibnamefont
  {Kitaev}},\ }\bibfield  {title} {\enquote {\bibinfo {title} {Anyons in an
  exactly solved model and beyond},}\ }\href {\doibase
  https://doi.org/10.1016/j.aop.2005.10.005} {\bibfield  {journal} {\bibinfo
  {journal} {Annals of Physics}\ }\textbf {\bibinfo {volume} {321}},\ \bibinfo
  {pages} {2 -- 111} (\bibinfo {year} {2006})},\ \bibinfo {note} {january
  Special Issue}\BibitemShut {NoStop}%
\bibitem [{\citenamefont {Jackeli}\ and\ \citenamefont
  {Khaliullin}(2009)}]{PhysRevLett.102.017205}%
  \BibitemOpen
  \bibfield  {author} {\bibinfo {author} {\bibfnamefont {G.}~\bibnamefont
  {Jackeli}}\ and\ \bibinfo {author} {\bibfnamefont {G.}~\bibnamefont
  {Khaliullin}},\ }\bibfield  {title} {\enquote {\bibinfo {title} {{Mott
  Insulators in the Strong Spin-Orbit Coupling Limit: From Heisenberg to a
  Quantum Compass and Kitaev Models}},}\ }\href {\doibase
  10.1103/PhysRevLett.102.017205} {\bibfield  {journal} {\bibinfo  {journal}
  {Phys. Rev. Lett.}\ }\textbf {\bibinfo {volume} {102}},\ \bibinfo {pages}
  {017205} (\bibinfo {year} {2009})}\BibitemShut {NoStop}%
\bibitem [{\citenamefont {Huang}\ \emph {et~al.}(2014)\citenamefont {Huang},
  \citenamefont {Chen},\ and\ \citenamefont {Hermele}}]{Huang2014}%
  \BibitemOpen
  \bibfield  {author} {\bibinfo {author} {\bibfnamefont {Yi-Ping}\ \bibnamefont
  {Huang}}, \bibinfo {author} {\bibfnamefont {Gang}\ \bibnamefont {Chen}}, \
  and\ \bibinfo {author} {\bibfnamefont {Michael}\ \bibnamefont {Hermele}},\
  }\bibfield  {title} {\enquote {\bibinfo {title} {{Quantum Spin Ices and
  Topological Phases from Dipolar-Octupolar Doublets on the Pyrochlore
  Lattice}},}\ }\href {\doibase 10.1103/PhysRevLett.112.167203} {\bibfield
  {journal} {\bibinfo  {journal} {Phys. Rev. Lett.}\ }\textbf {\bibinfo
  {volume} {112}},\ \bibinfo {pages} {167203} (\bibinfo {year}
  {2014})}\BibitemShut {NoStop}%
\bibitem [{\citenamefont {Li}\ \emph {et~al.}(2016)\citenamefont {Li},
  \citenamefont {Wang},\ and\ \citenamefont {Chen}}]{Yaodong2016}%
  \BibitemOpen
  \bibfield  {author} {\bibinfo {author} {\bibfnamefont {Yao-Dong}\
  \bibnamefont {Li}}, \bibinfo {author} {\bibfnamefont {Xiaoqun}\ \bibnamefont
  {Wang}}, \ and\ \bibinfo {author} {\bibfnamefont {Gang}\ \bibnamefont
  {Chen}},\ }\bibfield  {title} {\enquote {\bibinfo {title} {Hidden multipolar
  orders of dipole-octupole doublets on a triangular lattice},}\ }\href
  {\doibase 10.1103/PhysRevB.94.201114} {\bibfield  {journal} {\bibinfo
  {journal} {Phys. Rev. B}\ }\textbf {\bibinfo {volume} {94}},\ \bibinfo
  {pages} {201114} (\bibinfo {year} {2016})}\BibitemShut {NoStop}%
\bibitem [{\citenamefont {Hermele}\ \emph {et~al.}(2004)\citenamefont
  {Hermele}, \citenamefont {Fisher},\ and\ \citenamefont
  {Balents}}]{PhysRevB.69.064404}%
  \BibitemOpen
  \bibfield  {author} {\bibinfo {author} {\bibfnamefont {Michael}\ \bibnamefont
  {Hermele}}, \bibinfo {author} {\bibfnamefont {Matthew P.~A.}\ \bibnamefont
  {Fisher}}, \ and\ \bibinfo {author} {\bibfnamefont {Leon}\ \bibnamefont
  {Balents}},\ }\bibfield  {title} {\enquote {\bibinfo {title} {{Pyrochlore
  photons: The $U(1)$ spin liquid in a $S=\frac{1}{2}$ three-dimensional
  frustrated magnet}},}\ }\href {\doibase 10.1103/PhysRevB.69.064404}
  {\bibfield  {journal} {\bibinfo  {journal} {Phys. Rev. B}\ }\textbf {\bibinfo
  {volume} {69}},\ \bibinfo {pages} {064404} (\bibinfo {year}
  {2004})}\BibitemShut {NoStop}%
\bibitem [{\citenamefont {Henley}(2010)}]{Henley}%
  \BibitemOpen
  \bibfield  {author} {\bibinfo {author} {\bibfnamefont {Christopher~L.}\
  \bibnamefont {Henley}},\ }\bibfield  {title} {\enquote {\bibinfo {title}
  {{The “Coulomb Phase” in Frustrated Systems}},}\ }\href {\doibase
  10.1146/annurev-conmatphys-070909-104138} {\bibfield  {journal} {\bibinfo
  {journal} {Annual Review of Condensed Matter Physics}\ }\textbf {\bibinfo
  {volume} {1}},\ \bibinfo {pages} {179--210} (\bibinfo {year}
  {2010})}\BibitemShut {NoStop}%
\bibitem [{\citenamefont {Molavian}\ \emph {et~al.}(2007)\citenamefont
  {Molavian}, \citenamefont {Gingras},\ and\ \citenamefont
  {Canals}}]{PhysRevLett.98.157204}%
  \BibitemOpen
  \bibfield  {author} {\bibinfo {author} {\bibfnamefont {Hamid~R.}\
  \bibnamefont {Molavian}}, \bibinfo {author} {\bibfnamefont {Michel J.~P.}\
  \bibnamefont {Gingras}}, \ and\ \bibinfo {author} {\bibfnamefont {Benjamin}\
  \bibnamefont {Canals}},\ }\bibfield  {title} {\enquote {\bibinfo {title}
  {{Dynamically Induced Frustration as a Route to a Quantum Spin Ice State in
  ${\mathrm{Tb}}_{2}{\mathrm{Ti}}_{2}{\mathrm{O}}_{7}$ via Virtual Crystal
  Field Excitations and Quantum Many-Body Effects}},}\ }\href {\doibase
  10.1103/PhysRevLett.98.157204} {\bibfield  {journal} {\bibinfo  {journal}
  {Phys. Rev. Lett.}\ }\textbf {\bibinfo {volume} {98}},\ \bibinfo {pages}
  {157204} (\bibinfo {year} {2007})}\BibitemShut {NoStop}%
\bibitem [{\citenamefont {Ross}\ \emph {et~al.}(2011)\citenamefont {Ross},
  \citenamefont {Savary}, \citenamefont {Gaulin},\ and\ \citenamefont
  {Balents}}]{PhysRevX.1.021002}%
  \BibitemOpen
  \bibfield  {author} {\bibinfo {author} {\bibfnamefont {Kate~A.}\ \bibnamefont
  {Ross}}, \bibinfo {author} {\bibfnamefont {Lucile}\ \bibnamefont {Savary}},
  \bibinfo {author} {\bibfnamefont {Bruce~D.}\ \bibnamefont {Gaulin}}, \ and\
  \bibinfo {author} {\bibfnamefont {Leon}\ \bibnamefont {Balents}},\ }\bibfield
   {title} {\enquote {\bibinfo {title} {{Quantum Excitations in Quantum Spin
  Ice}},}\ }\href {\doibase 10.1103/PhysRevX.1.021002} {\bibfield  {journal}
  {\bibinfo  {journal} {Phys. Rev. X}\ }\textbf {\bibinfo {volume} {1}},\
  \bibinfo {pages} {021002} (\bibinfo {year} {2011})}\BibitemShut {NoStop}%
\bibitem [{\citenamefont {Hao}\ \emph {et~al.}(2014)\citenamefont {Hao},
  \citenamefont {Day},\ and\ \citenamefont {Gingras}}]{PhysRevB.90.214430}%
  \BibitemOpen
  \bibfield  {author} {\bibinfo {author} {\bibfnamefont {Zhihao}\ \bibnamefont
  {Hao}}, \bibinfo {author} {\bibfnamefont {Alexandre G.~R.}\ \bibnamefont
  {Day}}, \ and\ \bibinfo {author} {\bibfnamefont {Michel J.~P.}\ \bibnamefont
  {Gingras}},\ }\bibfield  {title} {\enquote {\bibinfo {title} {Bosonic
  many-body theory of quantum spin ice},}\ }\href {\doibase
  10.1103/PhysRevB.90.214430} {\bibfield  {journal} {\bibinfo  {journal} {Phys.
  Rev. B}\ }\textbf {\bibinfo {volume} {90}},\ \bibinfo {pages} {214430}
  (\bibinfo {year} {2014})}\BibitemShut {NoStop}%
\bibitem [{\citenamefont {Lee}\ \emph {et~al.}(2012)\citenamefont {Lee},
  \citenamefont {Onoda},\ and\ \citenamefont {Balents}}]{Sungbin2012}%
  \BibitemOpen
  \bibfield  {author} {\bibinfo {author} {\bibfnamefont {SungBin}\ \bibnamefont
  {Lee}}, \bibinfo {author} {\bibfnamefont {Shigeki}\ \bibnamefont {Onoda}}, \
  and\ \bibinfo {author} {\bibfnamefont {Leon}\ \bibnamefont {Balents}},\
  }\bibfield  {title} {\enquote {\bibinfo {title} {Generic quantum spin ice},}\
  }\href {\doibase 10.1103/PhysRevB.86.104412} {\bibfield  {journal} {\bibinfo
  {journal} {Phys. Rev. B}\ }\textbf {\bibinfo {volume} {86}},\ \bibinfo
  {pages} {104412} (\bibinfo {year} {2012})}\BibitemShut {NoStop}%
\bibitem [{\citenamefont {Savary}\ and\ \citenamefont
  {Balents}(2012)}]{PhysRevLett.108.037202}%
  \BibitemOpen
  \bibfield  {author} {\bibinfo {author} {\bibfnamefont {Lucile}\ \bibnamefont
  {Savary}}\ and\ \bibinfo {author} {\bibfnamefont {Leon}\ \bibnamefont
  {Balents}},\ }\bibfield  {title} {\enquote {\bibinfo {title} {{Coulombic
  Quantum Liquids in Spin-$1/2$ Pyrochlores}},}\ }\href {\doibase
  10.1103/PhysRevLett.108.037202} {\bibfield  {journal} {\bibinfo  {journal}
  {Phys. Rev. Lett.}\ }\textbf {\bibinfo {volume} {108}},\ \bibinfo {pages}
  {037202} (\bibinfo {year} {2012})}\BibitemShut {NoStop}%
\bibitem [{\citenamefont {Benton}\ \emph {et~al.}(2012)\citenamefont {Benton},
  \citenamefont {Sikora},\ and\ \citenamefont {Shannon}}]{PhysRevB.86.075154}%
  \BibitemOpen
  \bibfield  {author} {\bibinfo {author} {\bibfnamefont {Owen}\ \bibnamefont
  {Benton}}, \bibinfo {author} {\bibfnamefont {Olga}\ \bibnamefont {Sikora}}, \
  and\ \bibinfo {author} {\bibfnamefont {Nic}\ \bibnamefont {Shannon}},\
  }\bibfield  {title} {\enquote {\bibinfo {title} {Seeing the light:
  Experimental signatures of emergent electromagnetism in a quantum spin
  ice},}\ }\href {\doibase 10.1103/PhysRevB.86.075154} {\bibfield  {journal}
  {\bibinfo  {journal} {Phys. Rev. B}\ }\textbf {\bibinfo {volume} {86}},\
  \bibinfo {pages} {075154} (\bibinfo {year} {2012})}\BibitemShut {NoStop}%
\bibitem [{\citenamefont {Wan}\ and\ \citenamefont
  {Tchernyshyov}(2012)}]{PhysRevLett.108.247210}%
  \BibitemOpen
  \bibfield  {author} {\bibinfo {author} {\bibfnamefont {Yuan}\ \bibnamefont
  {Wan}}\ and\ \bibinfo {author} {\bibfnamefont {Oleg}\ \bibnamefont
  {Tchernyshyov}},\ }\bibfield  {title} {\enquote {\bibinfo {title} {{Quantum
  Strings in Quantum Spin Ice}},}\ }\href {\doibase
  10.1103/PhysRevLett.108.247210} {\bibfield  {journal} {\bibinfo  {journal}
  {Phys. Rev. Lett.}\ }\textbf {\bibinfo {volume} {108}},\ \bibinfo {pages}
  {247210} (\bibinfo {year} {2012})}\BibitemShut {NoStop}%
\bibitem [{\citenamefont {Sibille}\ \emph {et~al.}(2018)\citenamefont
  {Sibille}, \citenamefont {Gauthier}, \citenamefont {Yan}, \citenamefont
  {Hatnean}, \citenamefont {Ollivier}, \citenamefont {Winn}, \citenamefont
  {Filges}, \citenamefont {Balakrishnan}, \citenamefont {Kenzelmann},
  \citenamefont {Shannon},\ and\ \citenamefont {Fennell}}]{PrHfO}%
  \BibitemOpen
  \bibfield  {author} {\bibinfo {author} {\bibfnamefont {Romain}\ \bibnamefont
  {Sibille}}, \bibinfo {author} {\bibfnamefont {Nicolas}\ \bibnamefont
  {Gauthier}}, \bibinfo {author} {\bibfnamefont {Han}\ \bibnamefont {Yan}},
  \bibinfo {author} {\bibfnamefont {Monica~Ciomaga}\ \bibnamefont {Hatnean}},
  \bibinfo {author} {\bibfnamefont {Jacques}\ \bibnamefont {Ollivier}},
  \bibinfo {author} {\bibfnamefont {Barry}\ \bibnamefont {Winn}}, \bibinfo
  {author} {\bibfnamefont {Uwe}\ \bibnamefont {Filges}}, \bibinfo {author}
  {\bibfnamefont {Geetha}\ \bibnamefont {Balakrishnan}}, \bibinfo {author}
  {\bibfnamefont {Michel}\ \bibnamefont {Kenzelmann}}, \bibinfo {author}
  {\bibfnamefont {Nic}\ \bibnamefont {Shannon}}, \ and\ \bibinfo {author}
  {\bibfnamefont {Tom}\ \bibnamefont {Fennell}},\ }\bibfield  {title} {\enquote
  {\bibinfo {title} {{Experimental signatures of emergent quantum
  electrodynamics in Pr$_2$Hf$_2$O$_7$}},}\ }\href {\doibase
  10.1038/s41567-018-0116-x} {\bibfield  {journal} {\bibinfo  {journal} {Nature
  Physics}\ } (\bibinfo {year} {2018}),\ 10.1038/s41567-018-0116-x}\BibitemShut
  {NoStop}%
\bibitem [{\citenamefont {Chen}(2016)}]{PhysRevB.94.205107}%
  \BibitemOpen
  \bibfield  {author} {\bibinfo {author} {\bibfnamefont {Gang}\ \bibnamefont
  {Chen}},\ }\bibfield  {title} {\enquote {\bibinfo {title} {{``Magnetic
  monopole'' condensation of the pyrochlore ice U(1) quantum spin liquid:
  Application to ${\mathrm{Pr}}_{2}{\mathrm{Ir}}_{2}{\mathrm{O}}_{7}$ and
  ${\mathrm{Yb}}_{2}{\mathrm{Ti}}_{2}{\mathrm{O}}_{7}$}},}\ }\href {\doibase
  10.1103/PhysRevB.94.205107} {\bibfield  {journal} {\bibinfo  {journal} {Phys.
  Rev. B}\ }\textbf {\bibinfo {volume} {94}},\ \bibinfo {pages} {205107}
  (\bibinfo {year} {2016})}\BibitemShut {NoStop}%
\bibitem [{\citenamefont {Savary}\ and\ \citenamefont
  {Balents}(2017)}]{PhysRevLett.118.087203}%
  \BibitemOpen
  \bibfield  {author} {\bibinfo {author} {\bibfnamefont {Lucile}\ \bibnamefont
  {Savary}}\ and\ \bibinfo {author} {\bibfnamefont {Leon}\ \bibnamefont
  {Balents}},\ }\bibfield  {title} {\enquote {\bibinfo {title}
  {{Disorder-Induced Quantum Spin Liquid in Spin Ice Pyrochlores}},}\ }\href
  {\doibase 10.1103/PhysRevLett.118.087203} {\bibfield  {journal} {\bibinfo
  {journal} {Phys. Rev. Lett.}\ }\textbf {\bibinfo {volume} {118}},\ \bibinfo
  {pages} {087203} (\bibinfo {year} {2017})}\BibitemShut {NoStop}%
\bibitem [{\citenamefont {Wen}\ \emph {et~al.}(2017)\citenamefont {Wen},
  \citenamefont {Koohpayeh}, \citenamefont {Ross}, \citenamefont {Trump},
  \citenamefont {McQueen}, \citenamefont {Kimura}, \citenamefont {Nakatsuji},
  \citenamefont {Qiu}, \citenamefont {Pajerowski}, \citenamefont {Copley},\
  and\ \citenamefont {Broholm}}]{PhysRevLett.118.107206}%
  \BibitemOpen
  \bibfield  {author} {\bibinfo {author} {\bibfnamefont {J.-J.}\ \bibnamefont
  {Wen}}, \bibinfo {author} {\bibfnamefont {S.~M.}\ \bibnamefont {Koohpayeh}},
  \bibinfo {author} {\bibfnamefont {K.~A.}\ \bibnamefont {Ross}}, \bibinfo
  {author} {\bibfnamefont {B.~A.}\ \bibnamefont {Trump}}, \bibinfo {author}
  {\bibfnamefont {T.~M.}\ \bibnamefont {McQueen}}, \bibinfo {author}
  {\bibfnamefont {K.}~\bibnamefont {Kimura}}, \bibinfo {author} {\bibfnamefont
  {S.}~\bibnamefont {Nakatsuji}}, \bibinfo {author} {\bibfnamefont
  {Y.}~\bibnamefont {Qiu}}, \bibinfo {author} {\bibfnamefont {D.~M.}\
  \bibnamefont {Pajerowski}}, \bibinfo {author} {\bibfnamefont {J.~R.~D.}\
  \bibnamefont {Copley}}, \ and\ \bibinfo {author} {\bibfnamefont {C.~L.}\
  \bibnamefont {Broholm}},\ }\bibfield  {title} {\enquote {\bibinfo {title}
  {{Disordered Route to the Coulomb Quantum Spin Liquid: Random Transverse
  Fields on Spin Ice in
  ${\mathrm{Pr}}_{2}{\mathrm{Zr}}_{2}{\mathrm{O}}_{7}$}},}\ }\href {\doibase
  10.1103/PhysRevLett.118.107206} {\bibfield  {journal} {\bibinfo  {journal}
  {Phys. Rev. Lett.}\ }\textbf {\bibinfo {volume} {118}},\ \bibinfo {pages}
  {107206} (\bibinfo {year} {2017})}\BibitemShut {NoStop}%
\bibitem [{\citenamefont {Lantagne-Hurtubise}\ \emph
  {et~al.}(2017)\citenamefont {Lantagne-Hurtubise}, \citenamefont
  {Bhattacharjee},\ and\ \citenamefont {Moessner}}]{PhysRevB.96.125145}%
  \BibitemOpen
  \bibfield  {author} {\bibinfo {author} {\bibfnamefont {\'Etienne}\
  \bibnamefont {Lantagne-Hurtubise}}, \bibinfo {author} {\bibfnamefont
  {Subhro}\ \bibnamefont {Bhattacharjee}}, \ and\ \bibinfo {author}
  {\bibfnamefont {R.}~\bibnamefont {Moessner}},\ }\bibfield  {title} {\enquote
  {\bibinfo {title} {{Electric field control of emergent electrodynamics in
  quantum spin ice}},}\ }\href {\doibase 10.1103/PhysRevB.96.125145} {\bibfield
   {journal} {\bibinfo  {journal} {Phys. Rev. B}\ }\textbf {\bibinfo {volume}
  {96}},\ \bibinfo {pages} {125145} (\bibinfo {year} {2017})}\BibitemShut
  {NoStop}%
\bibitem [{\citenamefont {Chen}\ \emph {et~al.}(2014)\citenamefont {Chen},
  \citenamefont {Kee},\ and\ \citenamefont {Kim}}]{PhysRevLett.113.197202}%
  \BibitemOpen
  \bibfield  {author} {\bibinfo {author} {\bibfnamefont {Gang}\ \bibnamefont
  {Chen}}, \bibinfo {author} {\bibfnamefont {Hae-Young}\ \bibnamefont {Kee}}, \
  and\ \bibinfo {author} {\bibfnamefont {Yong~Baek}\ \bibnamefont {Kim}},\
  }\bibfield  {title} {\enquote {\bibinfo {title} {{Fractionalized Charge
  Excitations in a Spin Liquid on Partially Filled Pyrochlore Lattices}},}\
  }\href {\doibase 10.1103/PhysRevLett.113.197202} {\bibfield  {journal}
  {\bibinfo  {journal} {Phys. Rev. Lett.}\ }\textbf {\bibinfo {volume} {113}},\
  \bibinfo {pages} {197202} (\bibinfo {year} {2014})}\BibitemShut {NoStop}%
\bibitem [{\citenamefont {Chen}\ and\ \citenamefont
  {Lee}(2018)}]{PhysRevB.97.035124}%
  \BibitemOpen
  \bibfield  {author} {\bibinfo {author} {\bibfnamefont {Gang}\ \bibnamefont
  {Chen}}\ and\ \bibinfo {author} {\bibfnamefont {Patrick~A.}\ \bibnamefont
  {Lee}},\ }\bibfield  {title} {\enquote {\bibinfo {title} {{Emergent orbitals
  in the cluster Mott insulator on a breathing kagome lattice}},}\ }\href
  {\doibase 10.1103/PhysRevB.97.035124} {\bibfield  {journal} {\bibinfo
  {journal} {Phys. Rev. B}\ }\textbf {\bibinfo {volume} {97}},\ \bibinfo
  {pages} {035124} (\bibinfo {year} {2018})}\BibitemShut {NoStop}%
\bibitem [{\citenamefont {Chen}\ \emph {et~al.}(2016)\citenamefont {Chen},
  \citenamefont {Kee},\ and\ \citenamefont {Kim}}]{PhysRevB.93.245134}%
  \BibitemOpen
  \bibfield  {author} {\bibinfo {author} {\bibfnamefont {Gang}\ \bibnamefont
  {Chen}}, \bibinfo {author} {\bibfnamefont {Hae-Young}\ \bibnamefont {Kee}}, \
  and\ \bibinfo {author} {\bibfnamefont {Yong~Baek}\ \bibnamefont {Kim}},\
  }\bibfield  {title} {\enquote {\bibinfo {title} {{Cluster Mott insulators and
  two Curie-Weiss regimes on an anisotropic kagome lattice}},}\ }\href
  {\doibase 10.1103/PhysRevB.93.245134} {\bibfield  {journal} {\bibinfo
  {journal} {Phys. Rev. B}\ }\textbf {\bibinfo {volume} {93}},\ \bibinfo
  {pages} {245134} (\bibinfo {year} {2016})}\BibitemShut {NoStop}%
\bibitem [{\citenamefont {Lv}\ \emph {et~al.}(2015)\citenamefont {Lv},
  \citenamefont {Chen}, \citenamefont {Deng},\ and\ \citenamefont
  {Meng}}]{Chen2015}%
  \BibitemOpen
  \bibfield  {author} {\bibinfo {author} {\bibfnamefont {Jian-Ping}\
  \bibnamefont {Lv}}, \bibinfo {author} {\bibfnamefont {Gang}\ \bibnamefont
  {Chen}}, \bibinfo {author} {\bibfnamefont {Youjin}\ \bibnamefont {Deng}}, \
  and\ \bibinfo {author} {\bibfnamefont {Zi~Yang}\ \bibnamefont {Meng}},\
  }\bibfield  {title} {\enquote {\bibinfo {title} {{Coulomb Liquid Phases of
  Bosonic Cluster Mott Insulators on a Pyrochlore Lattice}},}\ }\href {\doibase
  10.1103/PhysRevLett.115.037202} {\bibfield  {journal} {\bibinfo  {journal}
  {Phys. Rev. Lett.}\ }\textbf {\bibinfo {volume} {115}},\ \bibinfo {pages}
  {037202} (\bibinfo {year} {2015})}\BibitemShut {NoStop}%
\bibitem [{\citenamefont {Carrasquilla}\ \emph {et~al.}(2017)\citenamefont
  {Carrasquilla}, \citenamefont {Chen},\ and\ \citenamefont
  {Melko}}]{PhysRevB.96.054405}%
  \BibitemOpen
  \bibfield  {author} {\bibinfo {author} {\bibfnamefont {Juan}\ \bibnamefont
  {Carrasquilla}}, \bibinfo {author} {\bibfnamefont {Gang}\ \bibnamefont
  {Chen}}, \ and\ \bibinfo {author} {\bibfnamefont {Roger~G.}\ \bibnamefont
  {Melko}},\ }\bibfield  {title} {\enquote {\bibinfo {title} {Tripartite
  entangled plaquette state in a cluster magnet},}\ }\href {\doibase
  10.1103/PhysRevB.96.054405} {\bibfield  {journal} {\bibinfo  {journal} {Phys.
  Rev. B}\ }\textbf {\bibinfo {volume} {96}},\ \bibinfo {pages} {054405}
  (\bibinfo {year} {2017})}\BibitemShut {NoStop}%
\bibitem [{\citenamefont {Carrasquilla}\ \emph {et~al.}(2015)\citenamefont
  {Carrasquilla}, \citenamefont {Hao},\ and\ \citenamefont
  {Melko}}]{Melko2015}%
  \BibitemOpen
  \bibfield  {author} {\bibinfo {author} {\bibfnamefont {Juan}\ \bibnamefont
  {Carrasquilla}}, \bibinfo {author} {\bibfnamefont {Zhihao}\ \bibnamefont
  {Hao}}, \ and\ \bibinfo {author} {\bibfnamefont {Roger}\ \bibnamefont
  {Melko}},\ }\bibfield  {title} {\enquote {\bibinfo {title} {A two-dimensional
  spin liquid in quantum kagome ice},}\ }\href {\doibase 10.1038/ncomms8421}
  {\bibfield  {journal} {\bibinfo  {journal} {Nature Communications}\ }\textbf
  {\bibinfo {volume} {6}},\ \bibinfo {pages} {7421} (\bibinfo {year}
  {2015})}\BibitemShut {NoStop}%
\bibitem [{\citenamefont {Hatnean}\ \emph {et~al.}(2015)\citenamefont
  {Hatnean}, \citenamefont {Lees}, \citenamefont {Petrenko}, \citenamefont
  {Keeble}, \citenamefont {Balakrishnan}, \citenamefont {Gutmann},
  \citenamefont {Klekovkina},\ and\ \citenamefont
  {Malkin}}]{PhysRevB.91.174416}%
  \BibitemOpen
  \bibfield  {author} {\bibinfo {author} {\bibfnamefont {M.~Ciomaga}\
  \bibnamefont {Hatnean}}, \bibinfo {author} {\bibfnamefont {M.~R.}\
  \bibnamefont {Lees}}, \bibinfo {author} {\bibfnamefont {O.~A.}\ \bibnamefont
  {Petrenko}}, \bibinfo {author} {\bibfnamefont {D.~S.}\ \bibnamefont
  {Keeble}}, \bibinfo {author} {\bibfnamefont {G.}~\bibnamefont
  {Balakrishnan}}, \bibinfo {author} {\bibfnamefont {M.~J.}\ \bibnamefont
  {Gutmann}}, \bibinfo {author} {\bibfnamefont {V.~V.}\ \bibnamefont
  {Klekovkina}}, \ and\ \bibinfo {author} {\bibfnamefont {B.~Z.}\ \bibnamefont
  {Malkin}},\ }\bibfield  {title} {\enquote {\bibinfo {title} {{Structural and
  magnetic investigations of single-crystalline neodymium zirconate pyrochlore
  ${\mathrm{Nd}}_{2}{\mathrm{Zr}}_{2}{\mathrm{O}}_{7}$}},}\ }\href {\doibase
  10.1103/PhysRevB.91.174416} {\bibfield  {journal} {\bibinfo  {journal} {Phys.
  Rev. B}\ }\textbf {\bibinfo {volume} {91}},\ \bibinfo {pages} {174416}
  (\bibinfo {year} {2015})}\BibitemShut {NoStop}%
\bibitem [{\citenamefont {Xu}\ \emph {et~al.}(2015)\citenamefont {Xu},
  \citenamefont {Anand}, \citenamefont {Bera}, \citenamefont {Frontzek},
  \citenamefont {Abernathy}, \citenamefont {Casati}, \citenamefont
  {Siemensmeyer},\ and\ \citenamefont {Lake}}]{PhysRevB.92.224430}%
  \BibitemOpen
  \bibfield  {author} {\bibinfo {author} {\bibfnamefont {J.}~\bibnamefont
  {Xu}}, \bibinfo {author} {\bibfnamefont {V.~K.}\ \bibnamefont {Anand}},
  \bibinfo {author} {\bibfnamefont {A.~K.}\ \bibnamefont {Bera}}, \bibinfo
  {author} {\bibfnamefont {M.}~\bibnamefont {Frontzek}}, \bibinfo {author}
  {\bibfnamefont {D.~L.}\ \bibnamefont {Abernathy}}, \bibinfo {author}
  {\bibfnamefont {N.}~\bibnamefont {Casati}}, \bibinfo {author} {\bibfnamefont
  {K.}~\bibnamefont {Siemensmeyer}}, \ and\ \bibinfo {author} {\bibfnamefont
  {B.}~\bibnamefont {Lake}},\ }\bibfield  {title} {\enquote {\bibinfo {title}
  {{Magnetic structure and crystal-field states of the pyrochlore
  antiferromagnet ${\mathrm{Nd}}_{2}{\mathrm{Zr}}_{2}{\mathrm{O}}_{7}$}},}\
  }\href {\doibase 10.1103/PhysRevB.92.224430} {\bibfield  {journal} {\bibinfo
  {journal} {Phys. Rev. B}\ }\textbf {\bibinfo {volume} {92}},\ \bibinfo
  {pages} {224430} (\bibinfo {year} {2015})}\BibitemShut {NoStop}%
\bibitem [{\citenamefont {Anand}\ \emph {et~al.}(2015)\citenamefont {Anand},
  \citenamefont {Bera}, \citenamefont {Xu}, \citenamefont {Herrmannsd\"orfer},
  \citenamefont {Ritter},\ and\ \citenamefont {Lake}}]{PhysRevB.92.184418}%
  \BibitemOpen
  \bibfield  {author} {\bibinfo {author} {\bibfnamefont {V.~K.}\ \bibnamefont
  {Anand}}, \bibinfo {author} {\bibfnamefont {A.~K.}\ \bibnamefont {Bera}},
  \bibinfo {author} {\bibfnamefont {J.}~\bibnamefont {Xu}}, \bibinfo {author}
  {\bibfnamefont {T.}~\bibnamefont {Herrmannsd\"orfer}}, \bibinfo {author}
  {\bibfnamefont {C.}~\bibnamefont {Ritter}}, \ and\ \bibinfo {author}
  {\bibfnamefont {B.}~\bibnamefont {Lake}},\ }\bibfield  {title} {\enquote
  {\bibinfo {title} {{Observation of long-range magnetic ordering in
  pyrohafnate ${\mathrm{Nd}}_{2}{\mathrm{Hf}}_{2}{\mathrm{O}}_{7}$: A neutron
  diffraction study}},}\ }\href {\doibase 10.1103/PhysRevB.92.184418}
  {\bibfield  {journal} {\bibinfo  {journal} {Phys. Rev. B}\ }\textbf {\bibinfo
  {volume} {92}},\ \bibinfo {pages} {184418} (\bibinfo {year}
  {2015})}\BibitemShut {NoStop}%
\bibitem [{\citenamefont {Bertin}\ \emph {et~al.}(2015)\citenamefont {Bertin},
  \citenamefont {Dalmas~de R\'eotier}, \citenamefont {F\aa{}k}, \citenamefont
  {Marin}, \citenamefont {Yaouanc}, \citenamefont {Forget}, \citenamefont
  {Sheptyakov}, \citenamefont {Frick}, \citenamefont {Ritter}, \citenamefont
  {Amato}, \citenamefont {Baines},\ and\ \citenamefont
  {King}}]{PhysRevB.92.144423}%
  \BibitemOpen
  \bibfield  {author} {\bibinfo {author} {\bibfnamefont {A.}~\bibnamefont
  {Bertin}}, \bibinfo {author} {\bibfnamefont {P.}~\bibnamefont {Dalmas~de
  R\'eotier}}, \bibinfo {author} {\bibfnamefont {B.}~\bibnamefont {F\aa{}k}},
  \bibinfo {author} {\bibfnamefont {C.}~\bibnamefont {Marin}}, \bibinfo
  {author} {\bibfnamefont {A.}~\bibnamefont {Yaouanc}}, \bibinfo {author}
  {\bibfnamefont {A.}~\bibnamefont {Forget}}, \bibinfo {author} {\bibfnamefont
  {D.}~\bibnamefont {Sheptyakov}}, \bibinfo {author} {\bibfnamefont
  {B.}~\bibnamefont {Frick}}, \bibinfo {author} {\bibfnamefont
  {C.}~\bibnamefont {Ritter}}, \bibinfo {author} {\bibfnamefont
  {A.}~\bibnamefont {Amato}}, \bibinfo {author} {\bibfnamefont
  {C.}~\bibnamefont {Baines}}, \ and\ \bibinfo {author} {\bibfnamefont
  {P.~J.~C.}\ \bibnamefont {King}},\ }\bibfield  {title} {\enquote {\bibinfo
  {title} {{${\mathrm{Nd}}_{2}{\mathrm{Sn}}_{2}{\mathrm{O}}_{7}$: An
  all-in--all-out pyrochlore magnet with no divergence-free field and
  anomalously slow paramagnetic spin dynamics}},}\ }\href {\doibase
  10.1103/PhysRevB.92.144423} {\bibfield  {journal} {\bibinfo  {journal} {Phys.
  Rev. B}\ }\textbf {\bibinfo {volume} {92}},\ \bibinfo {pages} {144423}
  (\bibinfo {year} {2015})}\BibitemShut {NoStop}%
\bibitem [{\citenamefont {Lhotel}\ \emph {et~al.}(2015)\citenamefont {Lhotel},
  \citenamefont {Petit}, \citenamefont {Guitteny}, \citenamefont {Florea},
  \citenamefont {Ciomaga~Hatnean}, \citenamefont {Colin}, \citenamefont
  {Ressouche}, \citenamefont {Lees},\ and\ \citenamefont
  {Balakrishnan}}]{PhysRevLett.115.197202}%
  \BibitemOpen
  \bibfield  {author} {\bibinfo {author} {\bibfnamefont {E.}~\bibnamefont
  {Lhotel}}, \bibinfo {author} {\bibfnamefont {S.}~\bibnamefont {Petit}},
  \bibinfo {author} {\bibfnamefont {S.}~\bibnamefont {Guitteny}}, \bibinfo
  {author} {\bibfnamefont {O.}~\bibnamefont {Florea}}, \bibinfo {author}
  {\bibfnamefont {M.}~\bibnamefont {Ciomaga~Hatnean}}, \bibinfo {author}
  {\bibfnamefont {C.}~\bibnamefont {Colin}}, \bibinfo {author} {\bibfnamefont
  {E.}~\bibnamefont {Ressouche}}, \bibinfo {author} {\bibfnamefont {M.~R.}\
  \bibnamefont {Lees}}, \ and\ \bibinfo {author} {\bibfnamefont
  {G.}~\bibnamefont {Balakrishnan}},\ }\bibfield  {title} {\enquote {\bibinfo
  {title} {{Fluctuations and All-In--All-Out Ordering in Dipole-Octupole
  ${\mathrm{Nd}}_{2}{\mathrm{Zr}}_{2}{\mathrm{O}}_{7}$}},}\ }\href {\doibase
  10.1103/PhysRevLett.115.197202} {\bibfield  {journal} {\bibinfo  {journal}
  {Phys. Rev. Lett.}\ }\textbf {\bibinfo {volume} {115}},\ \bibinfo {pages}
  {197202} (\bibinfo {year} {2015})}\BibitemShut {NoStop}%
\bibitem [{\citenamefont {Benton}(2016)}]{PhysRevB.94.104430}%
  \BibitemOpen
  \bibfield  {author} {\bibinfo {author} {\bibfnamefont {Owen}\ \bibnamefont
  {Benton}},\ }\bibfield  {title} {\enquote {\bibinfo {title} {{Quantum origins
  of moment fragmentation in
  ${\mathrm{Nd}}_{2}{\mathrm{Zr}}_{2}{\mathrm{O}}_{7}$}},}\ }\href {\doibase
  10.1103/PhysRevB.94.104430} {\bibfield  {journal} {\bibinfo  {journal} {Phys.
  Rev. B}\ }\textbf {\bibinfo {volume} {94}},\ \bibinfo {pages} {104430}
  (\bibinfo {year} {2016})}\BibitemShut {NoStop}%
\bibitem [{\citenamefont {Xu}\ \emph {et~al.}(2016)\citenamefont {Xu},
  \citenamefont {Balz}, \citenamefont {Baines}, \citenamefont {Luetkens},\ and\
  \citenamefont {Lake}}]{PhysRevB.94.064425}%
  \BibitemOpen
  \bibfield  {author} {\bibinfo {author} {\bibfnamefont {J.}~\bibnamefont
  {Xu}}, \bibinfo {author} {\bibfnamefont {C.}~\bibnamefont {Balz}}, \bibinfo
  {author} {\bibfnamefont {C.}~\bibnamefont {Baines}}, \bibinfo {author}
  {\bibfnamefont {H.}~\bibnamefont {Luetkens}}, \ and\ \bibinfo {author}
  {\bibfnamefont {B.}~\bibnamefont {Lake}},\ }\bibfield  {title} {\enquote
  {\bibinfo {title} {{Spin dynamics of the ordered dipolar-octupolar
  pseudospin-$\frac{1}{2}$ pyrochlore
  ${\text{Nd}}_{2}{\text{Zr}}_{2}{\text{O}}_{7}$ probed by muon spin
  relaxation}},}\ }\href {\doibase 10.1103/PhysRevB.94.064425} {\bibfield
  {journal} {\bibinfo  {journal} {Phys. Rev. B}\ }\textbf {\bibinfo {volume}
  {94}},\ \bibinfo {pages} {064425} (\bibinfo {year} {2016})}\BibitemShut
  {NoStop}%
\bibitem [{\citenamefont {Anand}\ \emph {et~al.}(2017)\citenamefont {Anand},
  \citenamefont {Abernathy}, \citenamefont {Adroja}, \citenamefont {Hillier},
  \citenamefont {Biswas},\ and\ \citenamefont {Lake}}]{PhysRevB.95.224420}%
  \BibitemOpen
  \bibfield  {author} {\bibinfo {author} {\bibfnamefont {V.~K.}\ \bibnamefont
  {Anand}}, \bibinfo {author} {\bibfnamefont {D.~L.}\ \bibnamefont
  {Abernathy}}, \bibinfo {author} {\bibfnamefont {D.~T.}\ \bibnamefont
  {Adroja}}, \bibinfo {author} {\bibfnamefont {A.~D.}\ \bibnamefont {Hillier}},
  \bibinfo {author} {\bibfnamefont {P.~K.}\ \bibnamefont {Biswas}}, \ and\
  \bibinfo {author} {\bibfnamefont {B.}~\bibnamefont {Lake}},\ }\bibfield
  {title} {\enquote {\bibinfo {title} {{Muon spin relaxation and inelastic
  neutron scattering investigations of the all-in/all-out antiferromagnet
  ${\mathrm{Nd}}_{2}{\mathrm{Hf}}_{2}{\mathrm{O}}_{7}$}},}\ }\href {\doibase
  10.1103/PhysRevB.95.224420} {\bibfield  {journal} {\bibinfo  {journal} {Phys.
  Rev. B}\ }\textbf {\bibinfo {volume} {95}},\ \bibinfo {pages} {224420}
  (\bibinfo {year} {2017})}\BibitemShut {NoStop}%
\bibitem [{\citenamefont {Dalmas~de R\'eotier}\ \emph
  {et~al.}(2017)\citenamefont {Dalmas~de R\'eotier}, \citenamefont {Yaouanc},
  \citenamefont {Maisuradze}, \citenamefont {Bertin}, \citenamefont {Baker},
  \citenamefont {Hillier},\ and\ \citenamefont {Forget}}]{PhysRevB.95.134420}%
  \BibitemOpen
  \bibfield  {author} {\bibinfo {author} {\bibfnamefont {P.}~\bibnamefont
  {Dalmas~de R\'eotier}}, \bibinfo {author} {\bibfnamefont {A.}~\bibnamefont
  {Yaouanc}}, \bibinfo {author} {\bibfnamefont {A.}~\bibnamefont {Maisuradze}},
  \bibinfo {author} {\bibfnamefont {A.}~\bibnamefont {Bertin}}, \bibinfo
  {author} {\bibfnamefont {P.~J.}\ \bibnamefont {Baker}}, \bibinfo {author}
  {\bibfnamefont {A.~D.}\ \bibnamefont {Hillier}}, \ and\ \bibinfo {author}
  {\bibfnamefont {A.}~\bibnamefont {Forget}},\ }\bibfield  {title} {\enquote
  {\bibinfo {title} {{Slow spin tunneling in the paramagnetic phase of the
  pyrochlore ${\mathrm{Nd}}_{2}{\mathrm{Sn}}_{2}{\mathrm{O}}_{7}$}},}\ }\href
  {\doibase 10.1103/PhysRevB.95.134420} {\bibfield  {journal} {\bibinfo
  {journal} {Phys. Rev. B}\ }\textbf {\bibinfo {volume} {95}},\ \bibinfo
  {pages} {134420} (\bibinfo {year} {2017})}\BibitemShut {NoStop}%
\bibitem [{\citenamefont {Li}\ and\ \citenamefont
  {Chen}(2017)}]{PhysRevB.95.041106}%
  \BibitemOpen
  \bibfield  {author} {\bibinfo {author} {\bibfnamefont {Yao-Dong}\
  \bibnamefont {Li}}\ and\ \bibinfo {author} {\bibfnamefont {Gang}\
  \bibnamefont {Chen}},\ }\bibfield  {title} {\enquote {\bibinfo {title}
  {{Symmetry enriched U(1) topological orders for dipole-octupole doublets on a
  pyrochlore lattice}},}\ }\href {\doibase 10.1103/PhysRevB.95.041106}
  {\bibfield  {journal} {\bibinfo  {journal} {Phys. Rev. B}\ }\textbf {\bibinfo
  {volume} {95}},\ \bibinfo {pages} {041106} (\bibinfo {year}
  {2017})}\BibitemShut {NoStop}%
\bibitem [{\citenamefont {Sibille}\ \emph {et~al.}(2015)\citenamefont
  {Sibille}, \citenamefont {Lhotel}, \citenamefont {Pomjakushin}, \citenamefont
  {Baines}, \citenamefont {Fennell},\ and\ \citenamefont
  {Kenzelmann}}]{PhysRevLett.115.097202}%
  \BibitemOpen
  \bibfield  {author} {\bibinfo {author} {\bibfnamefont {Romain}\ \bibnamefont
  {Sibille}}, \bibinfo {author} {\bibfnamefont {Elsa}\ \bibnamefont {Lhotel}},
  \bibinfo {author} {\bibfnamefont {Vladimir}\ \bibnamefont {Pomjakushin}},
  \bibinfo {author} {\bibfnamefont {Chris}\ \bibnamefont {Baines}}, \bibinfo
  {author} {\bibfnamefont {Tom}\ \bibnamefont {Fennell}}, \ and\ \bibinfo
  {author} {\bibfnamefont {Michel}\ \bibnamefont {Kenzelmann}},\ }\bibfield
  {title} {\enquote {\bibinfo {title} {{Candidate Quantum Spin Liquid in the
  ${\mathrm{Ce}}^{3+}$ Pyrochlore Stannate
  ${\mathrm{Ce}}_{2}{\mathrm{Sn}}_{2}{\mathrm{O}}_{7}$}},}\ }\href {\doibase
  10.1103/PhysRevLett.115.097202} {\bibfield  {journal} {\bibinfo  {journal}
  {Phys. Rev. Lett.}\ }\textbf {\bibinfo {volume} {115}},\ \bibinfo {pages}
  {097202} (\bibinfo {year} {2015})}\BibitemShut {NoStop}%
\bibitem [{\citenamefont {Banerjee}\ \emph {et~al.}(2008)\citenamefont
  {Banerjee}, \citenamefont {Isakov}, \citenamefont {Damle},\ and\
  \citenamefont {Kim}}]{PhysRevLett.100.047208}%
  \BibitemOpen
  \bibfield  {author} {\bibinfo {author} {\bibfnamefont {Argha}\ \bibnamefont
  {Banerjee}}, \bibinfo {author} {\bibfnamefont {Sergei~V.}\ \bibnamefont
  {Isakov}}, \bibinfo {author} {\bibfnamefont {Kedar}\ \bibnamefont {Damle}}, \
  and\ \bibinfo {author} {\bibfnamefont {Yong~Baek}\ \bibnamefont {Kim}},\
  }\bibfield  {title} {\enquote {\bibinfo {title} {{Unusual Liquid State of
  Hard-Core Bosons on the Pyrochlore Lattice}},}\ }\href {\doibase
  10.1103/PhysRevLett.100.047208} {\bibfield  {journal} {\bibinfo  {journal}
  {Phys. Rev. Lett.}\ }\textbf {\bibinfo {volume} {100}},\ \bibinfo {pages}
  {047208} (\bibinfo {year} {2008})}\BibitemShut {NoStop}%
\bibitem [{\citenamefont {Onoda}\ and\ \citenamefont
  {Tanaka}(2010)}]{PhysRevLett.105.047201}%
  \BibitemOpen
  \bibfield  {author} {\bibinfo {author} {\bibfnamefont {Shigeki}\ \bibnamefont
  {Onoda}}\ and\ \bibinfo {author} {\bibfnamefont {Yoichi}\ \bibnamefont
  {Tanaka}},\ }\bibfield  {title} {\enquote {\bibinfo {title} {{Quantum Melting
  of Spin Ice: Emergent Cooperative Quadrupole and Chirality}},}\ }\href
  {\doibase 10.1103/PhysRevLett.105.047201} {\bibfield  {journal} {\bibinfo
  {journal} {Phys. Rev. Lett.}\ }\textbf {\bibinfo {volume} {105}},\ \bibinfo
  {pages} {047201} (\bibinfo {year} {2010})}\BibitemShut {NoStop}%
\bibitem [{\citenamefont {Shannon}\ \emph {et~al.}(2012)\citenamefont
  {Shannon}, \citenamefont {Sikora}, \citenamefont {Pollmann}, \citenamefont
  {Penc},\ and\ \citenamefont {Fulde}}]{PhysRevLett.108.067204}%
  \BibitemOpen
  \bibfield  {author} {\bibinfo {author} {\bibfnamefont {Nic}\ \bibnamefont
  {Shannon}}, \bibinfo {author} {\bibfnamefont {Olga}\ \bibnamefont {Sikora}},
  \bibinfo {author} {\bibfnamefont {Frank}\ \bibnamefont {Pollmann}}, \bibinfo
  {author} {\bibfnamefont {Karlo}\ \bibnamefont {Penc}}, \ and\ \bibinfo
  {author} {\bibfnamefont {Peter}\ \bibnamefont {Fulde}},\ }\bibfield  {title}
  {\enquote {\bibinfo {title} {Quantum ice: A quantum monte carlo study},}\
  }\href {\doibase 10.1103/PhysRevLett.108.067204} {\bibfield  {journal}
  {\bibinfo  {journal} {Phys. Rev. Lett.}\ }\textbf {\bibinfo {volume} {108}},\
  \bibinfo {pages} {067204} (\bibinfo {year} {2012})}\BibitemShut {NoStop}%
\bibitem [{\citenamefont {Kato}\ and\ \citenamefont
  {Onoda}(2015)}]{PhysRevLett.115.077202}%
  \BibitemOpen
  \bibfield  {author} {\bibinfo {author} {\bibfnamefont {Yasuyuki}\
  \bibnamefont {Kato}}\ and\ \bibinfo {author} {\bibfnamefont {Shigeki}\
  \bibnamefont {Onoda}},\ }\bibfield  {title} {\enquote {\bibinfo {title}
  {{Numerical Evidence of Quantum Melting of Spin Ice: Quantum-to-Classical
  Crossover}},}\ }\href {\doibase 10.1103/PhysRevLett.115.077202} {\bibfield
  {journal} {\bibinfo  {journal} {Phys. Rev. Lett.}\ }\textbf {\bibinfo
  {volume} {115}},\ \bibinfo {pages} {077202} (\bibinfo {year}
  {2015})}\BibitemShut {NoStop}%
\bibitem [{\citenamefont {Glaetzle}\ \emph {et~al.}(2015)\citenamefont
  {Glaetzle}, \citenamefont {Dalmonte}, \citenamefont {Nath}, \citenamefont
  {Gross}, \citenamefont {Bloch},\ and\ \citenamefont
  {Zoller}}]{PhysRevLett.114.173002}%
  \BibitemOpen
  \bibfield  {author} {\bibinfo {author} {\bibfnamefont {Alexander~W.}\
  \bibnamefont {Glaetzle}}, \bibinfo {author} {\bibfnamefont {Marcello}\
  \bibnamefont {Dalmonte}}, \bibinfo {author} {\bibfnamefont {Rejish}\
  \bibnamefont {Nath}}, \bibinfo {author} {\bibfnamefont {Christian}\
  \bibnamefont {Gross}}, \bibinfo {author} {\bibfnamefont {Immanuel}\
  \bibnamefont {Bloch}}, \ and\ \bibinfo {author} {\bibfnamefont {Peter}\
  \bibnamefont {Zoller}},\ }\bibfield  {title} {\enquote {\bibinfo {title}
  {{Designing Frustrated Quantum Magnets with Laser-Dressed Rydberg Atoms}},}\
  }\href {\doibase 10.1103/PhysRevLett.114.173002} {\bibfield  {journal}
  {\bibinfo  {journal} {Phys. Rev. Lett.}\ }\textbf {\bibinfo {volume} {114}},\
  \bibinfo {pages} {173002} (\bibinfo {year} {2015})}\BibitemShut {NoStop}%
\bibitem [{\citenamefont {B\"uchler}\ \emph {et~al.}(2005)\citenamefont
  {B\"uchler}, \citenamefont {Hermele}, \citenamefont {Huber}, \citenamefont
  {Fisher},\ and\ \citenamefont {Zoller}}]{PhysRevLett.95.040402}%
  \BibitemOpen
  \bibfield  {author} {\bibinfo {author} {\bibfnamefont {H.~P.}\ \bibnamefont
  {B\"uchler}}, \bibinfo {author} {\bibfnamefont {M.}~\bibnamefont {Hermele}},
  \bibinfo {author} {\bibfnamefont {S.~D.}\ \bibnamefont {Huber}}, \bibinfo
  {author} {\bibfnamefont {Matthew P.~A.}\ \bibnamefont {Fisher}}, \ and\
  \bibinfo {author} {\bibfnamefont {P.}~\bibnamefont {Zoller}},\ }\bibfield
  {title} {\enquote {\bibinfo {title} {{Atomic Quantum Simulator for Lattice
  Gauge Theories and Ring Exchange Models}},}\ }\href {\doibase
  10.1103/PhysRevLett.95.040402} {\bibfield  {journal} {\bibinfo  {journal}
  {Phys. Rev. Lett.}\ }\textbf {\bibinfo {volume} {95}},\ \bibinfo {pages}
  {040402} (\bibinfo {year} {2005})}\BibitemShut {NoStop}%
\bibitem [{\citenamefont {Bramwell}\ and\ \citenamefont
  {Gingras}(2001)}]{Bramwell2001}%
  \BibitemOpen
  \bibfield  {author} {\bibinfo {author} {\bibfnamefont {S.~T.}\ \bibnamefont
  {Bramwell}}\ and\ \bibinfo {author} {\bibfnamefont {M.J.P}\ \bibnamefont
  {Gingras}},\ }\bibfield  {title} {\enquote {\bibinfo {title} {{Spin Ice State
  in Frustrated Magnetic Pyrochlore Materials}},}\ }\href {\doibase
  10.1126/science.1064761} {\bibfield  {journal} {\bibinfo  {journal}
  {Science}\ }\textbf {\bibinfo {volume} {294}},\ \bibinfo {pages} {1495--1501}
  (\bibinfo {year} {2001})}\BibitemShut {NoStop}%
\bibitem [{\citenamefont {Castelnovo}\ \emph {et~al.}(2008)\citenamefont
  {Castelnovo}, \citenamefont {Moessner},\ and\ \citenamefont
  {Sondhi}}]{Castelnovo2008}%
  \BibitemOpen
  \bibfield  {author} {\bibinfo {author} {\bibfnamefont {C.}~\bibnamefont
  {Castelnovo}}, \bibinfo {author} {\bibfnamefont {R.}~\bibnamefont
  {Moessner}}, \ and\ \bibinfo {author} {\bibfnamefont {S.~L.}\ \bibnamefont
  {Sondhi}},\ }\bibfield  {title} {\enquote {\bibinfo {title} {{Magnetic
  monopoles in spin ice}},}\ }\href@noop {} {\bibfield  {journal} {\bibinfo
  {journal} {Nature}\ }\textbf {\bibinfo {volume} {451}},\ \bibinfo {pages}
  {42--45} (\bibinfo {year} {2008})}\BibitemShut {NoStop}%
\bibitem [{\citenamefont {Gingras}\ and\ \citenamefont
  {McClarty}(2014)}]{0034-4885-77-5-056501}%
  \BibitemOpen
  \bibfield  {author} {\bibinfo {author} {\bibfnamefont {M~J~P}\ \bibnamefont
  {Gingras}}\ and\ \bibinfo {author} {\bibfnamefont {P~A}\ \bibnamefont
  {McClarty}},\ }\bibfield  {title} {\enquote {\bibinfo {title} {{Quantum spin
  ice: a search for gapless quantum spin liquids in pyrochlore magnets}},}\
  }\href {http://stacks.iop.org/0034-4885/77/i=5/a=056501} {\bibfield
  {journal} {\bibinfo  {journal} {Reports on Progress in Physics}\ }\textbf
  {\bibinfo {volume} {77}},\ \bibinfo {pages} {056501} (\bibinfo {year}
  {2014})}\BibitemShut {NoStop}%
\bibitem [{\citenamefont {Prokof'ev}\ \emph
  {et~al.}(1998{\natexlab{a}})\citenamefont {Prokof'ev}, \citenamefont
  {Svistunov},\ and\ \citenamefont {Tupitsyn}}]{Prokofev1998}%
  \BibitemOpen
  \bibfield  {author} {\bibinfo {author} {\bibfnamefont {N.~V.}\ \bibnamefont
  {Prokof'ev}}, \bibinfo {author} {\bibfnamefont {B.~V.}\ \bibnamefont
  {Svistunov}}, \ and\ \bibinfo {author} {\bibfnamefont {I.~S.}\ \bibnamefont
  {Tupitsyn}},\ }\bibfield  {title} {\enquote {\bibinfo {title} {Exact,
  complete, and universal continuous-time worldline monte carlo approach to the
  statistics of discrete quantum systems},}\ }\href {\doibase 10.1134/1.558661}
  {\bibfield  {journal} {\bibinfo  {journal} {Journal of Experimental and
  Theoretical Physics}\ }\textbf {\bibinfo {volume} {87}},\ \bibinfo {pages}
  {310--321} (\bibinfo {year} {1998}{\natexlab{a}})}\BibitemShut {NoStop}%
\bibitem [{\citenamefont {Prokof'ev}\ \emph
  {et~al.}(1998{\natexlab{b}})\citenamefont {Prokof'ev}, \citenamefont
  {Svistunov},\ and\ \citenamefont {Tupitsyn}}]{Prokofev1998b}%
  \BibitemOpen
  \bibfield  {author} {\bibinfo {author} {\bibfnamefont {N.V}\ \bibnamefont
  {Prokof'ev}}, \bibinfo {author} {\bibfnamefont {B.V}\ \bibnamefont
  {Svistunov}}, \ and\ \bibinfo {author} {\bibfnamefont {I.S}\ \bibnamefont
  {Tupitsyn}},\ }\bibfield  {title} {\enquote {\bibinfo {title} {“worm”
  algorithm in quantum monte carlo simulations},}\ }\href {\doibase
  https://doi.org/10.1016/S0375-9601(97)00957-2} {\bibfield  {journal}
  {\bibinfo  {journal} {Physics Letters A}\ }\textbf {\bibinfo {volume}
  {238}},\ \bibinfo {pages} {253 -- 257} (\bibinfo {year}
  {1998}{\natexlab{b}})}\BibitemShut {NoStop}%
\bibitem [{Note1()}]{Note1}%
  \BibitemOpen
  \bibinfo {note} {It was recently realized in Ref.~\protect \rev@citealpnum
  {PhysRevB.96.195127} that at higher energies, the density correlator would
  include the magnetic monopole contribution.}\BibitemShut {Stop}%
\bibitem [{\citenamefont {Chen}(2017{\natexlab{a}})}]{PhysRevB.96.085136}%
  \BibitemOpen
  \bibfield  {author} {\bibinfo {author} {\bibfnamefont {Gang}\ \bibnamefont
  {Chen}},\ }\bibfield  {title} {\enquote {\bibinfo {title} {Spectral
  periodicity of the spinon continuum in quantum spin ice},}\ }\href {\doibase
  10.1103/PhysRevB.96.085136} {\bibfield  {journal} {\bibinfo  {journal} {Phys.
  Rev. B}\ }\textbf {\bibinfo {volume} {96}},\ \bibinfo {pages} {085136}
  (\bibinfo {year} {2017}{\natexlab{a}})}\BibitemShut {NoStop}%
\bibitem [{\citenamefont {Chen}(2017{\natexlab{b}})}]{PhysRevB.96.195127}%
  \BibitemOpen
  \bibfield  {author} {\bibinfo {author} {\bibfnamefont {Gang}\ \bibnamefont
  {Chen}},\ }\bibfield  {title} {\enquote {\bibinfo {title} {{Dirac's
  ``magnetic monopoles'' in pyrochlore ice $U(1)$ spin liquids: Spectrum and
  classification}},}\ }\href {\doibase 10.1103/PhysRevB.96.195127} {\bibfield
  {journal} {\bibinfo  {journal} {Phys. Rev. B}\ }\textbf {\bibinfo {volume}
  {96}},\ \bibinfo {pages} {195127} (\bibinfo {year}
  {2017}{\natexlab{b}})}\BibitemShut {NoStop}%
\bibitem [{\citenamefont {Mauws}\ \emph {et~al.}(2018)\citenamefont {Mauws},
  \citenamefont {Hallas}, \citenamefont {Sala}, \citenamefont {Aczel},
  \citenamefont {Sarte}, \citenamefont {Gaudet}, \citenamefont {Ziat},
  \citenamefont {Quilliam}, \citenamefont {Lussier}, \citenamefont {Bieringer},
  \citenamefont {Zhou}, \citenamefont {Wildes}, \citenamefont {Stone},
  \citenamefont {Abernathy}, \citenamefont {Luke}, \citenamefont {Gaulin},\
  and\ \citenamefont {Wiebe}}]{SmTiO}%
  \BibitemOpen
  \bibfield  {author} {\bibinfo {author} {\bibfnamefont {C.}~\bibnamefont
  {Mauws}}, \bibinfo {author} {\bibfnamefont {A.~M.}\ \bibnamefont {Hallas}},
  \bibinfo {author} {\bibfnamefont {G.}~\bibnamefont {Sala}}, \bibinfo {author}
  {\bibfnamefont {A.~A.}\ \bibnamefont {Aczel}}, \bibinfo {author}
  {\bibfnamefont {P.~M.}\ \bibnamefont {Sarte}}, \bibinfo {author}
  {\bibfnamefont {J.}~\bibnamefont {Gaudet}}, \bibinfo {author} {\bibfnamefont
  {D.}~\bibnamefont {Ziat}}, \bibinfo {author} {\bibfnamefont {J.~A.}\
  \bibnamefont {Quilliam}}, \bibinfo {author} {\bibfnamefont {J.~A.}\
  \bibnamefont {Lussier}}, \bibinfo {author} {\bibfnamefont {M.}~\bibnamefont
  {Bieringer}}, \bibinfo {author} {\bibfnamefont {H.~D.}\ \bibnamefont {Zhou}},
  \bibinfo {author} {\bibfnamefont {A.}~\bibnamefont {Wildes}}, \bibinfo
  {author} {\bibfnamefont {M.~B.}\ \bibnamefont {Stone}}, \bibinfo {author}
  {\bibfnamefont {D.}~\bibnamefont {Abernathy}}, \bibinfo {author}
  {\bibfnamefont {G.~M.}\ \bibnamefont {Luke}}, \bibinfo {author}
  {\bibfnamefont {B.~D.}\ \bibnamefont {Gaulin}}, \ and\ \bibinfo {author}
  {\bibfnamefont {C.~R.}\ \bibnamefont {Wiebe}},\ }\bibfield  {title} {\enquote
  {\bibinfo {title} {{Dipolar-Octupolar Ising Antiferromagnetism in
  Sm$_2$Ti$_2$O$_7$: A Moment Fragmentation Candidate}},}\ }\href@noop {}
  {\bibfield  {journal} {\bibinfo  {journal} {arXiv}\ }\textbf {\bibinfo
  {volume} {1805.09472}} (\bibinfo {year} {2018})}\BibitemShut {NoStop}%
\bibitem [{\citenamefont {Gardner}\ \emph {et~al.}(2010)\citenamefont
  {Gardner}, \citenamefont {Gingras},\ and\ \citenamefont
  {Greedan}}]{RevModPhys.82.53}%
  \BibitemOpen
  \bibfield  {author} {\bibinfo {author} {\bibfnamefont {Jason~S.}\
  \bibnamefont {Gardner}}, \bibinfo {author} {\bibfnamefont {Michel J.~P.}\
  \bibnamefont {Gingras}}, \ and\ \bibinfo {author} {\bibfnamefont {John~E.}\
  \bibnamefont {Greedan}},\ }\bibfield  {title} {\enquote {\bibinfo {title}
  {Magnetic pyrochlore oxides},}\ }\href {\doibase 10.1103/RevModPhys.82.53}
  {\bibfield  {journal} {\bibinfo  {journal} {Rev. Mod. Phys.}\ }\textbf
  {\bibinfo {volume} {82}},\ \bibinfo {pages} {53--107} (\bibinfo {year}
  {2010})}\BibitemShut {NoStop}%
\end{thebibliography}%

\vspace{1cm}

\noindent{\emph{\bf Acknowledgments}}\\
{\footnotesize \noindent{G.C.} would like to thank the hospitality 
of Prof Zhong Wang from IAS Tsinghua where this work is completed. 
G.C. acknowledges a previous collaboration with Mike Hermele from 
University of Colorado Boulder and Professor Xiaoqun Wang for the 
many support and encouragements. This work is supported by 
the National Natural Science Foundation of China under Grants 
No.11625522 (CJH, YJD), the Ministry 
of Science and Technology of China No.2016YFA0301604 (CJH, YJD),
and No.2016YFA0301001 (CLL, GC), the Start-Up Funds and the Program 
of First-Class Construction of Fudan University (CLL, GC), and the 
Thousand-Youth-Talent Program (CLL, GC) of China.}
\\

\noindent{\emph{\bf Contributions}}\\
{\footnotesize \noindent{Gang Chen} designed and planned the whole project. 
Gang Chen wrote this manuscript with the algorithm input from Chun-Jiong Huang
and extensive discussion with Changle Liu and Chun-Jiong Huang.
Gang Chen and Chun-Jiong Huang wrote the Supplementary material. 
Chun-Jiong Huang and Youjin Deng designed the Worm-type quantum Monte 
Carlo codes and performed the numerical simulation on the clusters of 
Youjin Deng's group at University of Science and Technology of China. 
Changle Liu and Gang Chen carried out the theoretical analysis and 
discussion. Chun-Jiong Huang, Changle Liu and Gang Chen
analysed the data. All authors commented on the results.}
\\

\noindent {\bf Additional information}

{\footnotesize\noindent Correspondence and requests for materials should be 
addressed to G.C. (gangchen.physics@gmail.com) or Y.J.D. (yjdeng@ustc.edu.cn).} \\ 

{\footnotesize \noindent  {\bf Competing financial interests:}
The authors declare no competing financial interests.}

\vspace{2cm}
\newpage

{\Large {Supplementary Materials for ``Extended Coulomb 
liquid of paired hardcore boson model on a pyrochlore lattice''}}
\vspace{1cm}

\noindent{\bf Perturbation theory.} For the completeness, we provide a 
perturbative analysis and understanding of our paired hardcore boson model. 
In the well-known limit without boson pairing ({\sl i.e.} ${t_2 =0}$), the 
third-order degenerate perturbation within the spin ice manifold 
generates a three-boson collective hopping on the elementary hexagon 
of the pyochlore lattice that is given by~\cite{PhysRevB.69.064404}
\begin{eqnarray}
H_{\text{eff}} = -t_{\text{coll}} \sum_{\hexagon}
[ b^\dagger_1 b^{}_2 b^\dagger_3 b^{}_4 b^{\dagger}_5 b^{}_6 + h.c.],
\end{eqnarray}
where ${t_{\text{coll}} = 12t_1^3/V^2}$ is positive for ${t_1>0}$,
and $1,2,3,4,5,6$ are the six lattice sites on the perimeter of the hexagon. 
In the opposite case with $t_1=0$ and $t_2\neq 0$, we need a
six order perturbation theory. Within this low-energy manifold, 
one then expresses the hardcore bosons as $b^\dagger_i 
\sim e^{i A_{{\boldsymbol r}{\boldsymbol r}'}}$ 
where $A_{{\boldsymbol r}{\boldsymbol r}'}$ is the $U(1)$ vector gauge 
potential on the link connecting the centers of neighboring tetrahedra, 
and the effective Hamiltonian simply becomes~\cite{PhysRevB.69.064404} 
\begin{eqnarray}
H_{\text{eff}} = -2t_{\text{coll}} \sum_{\hexagon^{\ast}} \cos ( curl A ),
\end{eqnarray}
where a positive $t_{\text{coll}}$ favors a zero-flux sector with ${curl A=0}$
for the spinons
and $\hexagon^{\ast}$ refers to the elementary hexagon on the diamond lattice 
formed by the tetrahedral centers.

\begin{figure}[h]
\includegraphics[width=0.8\columnwidth]{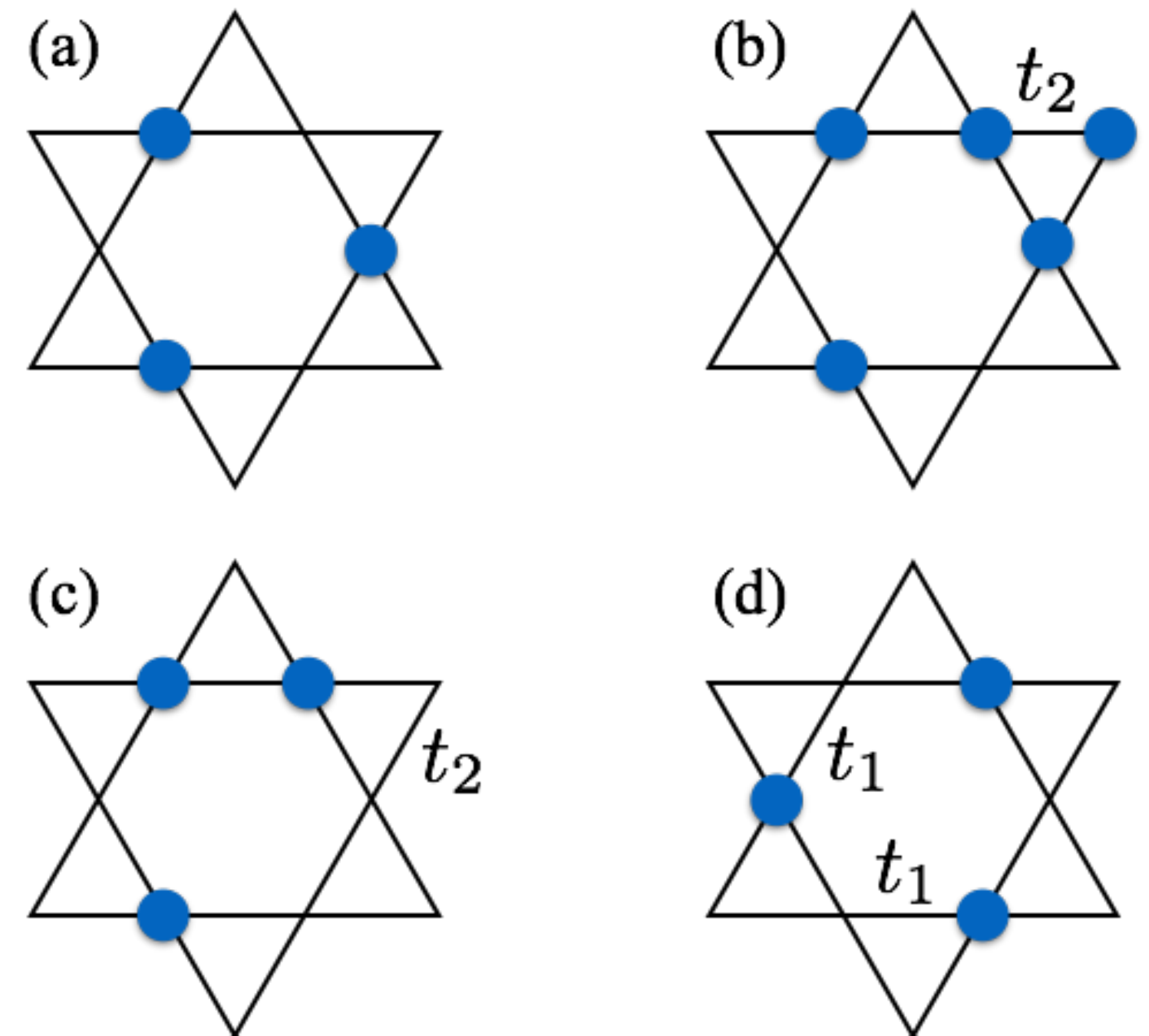}
\caption{{\bf The fourth order perturbation.} 
Here we apply two ``$t_2$''pairing processes and two $t_1$ hoppings.
The location of $t_1$ or $t_2$ indicates the bond that the $t_1$ hopping
or $t_2$ pairing is applied. }
\label{hop}
\end{figure}

\begin{figure}[h]
\includegraphics[width=\columnwidth]{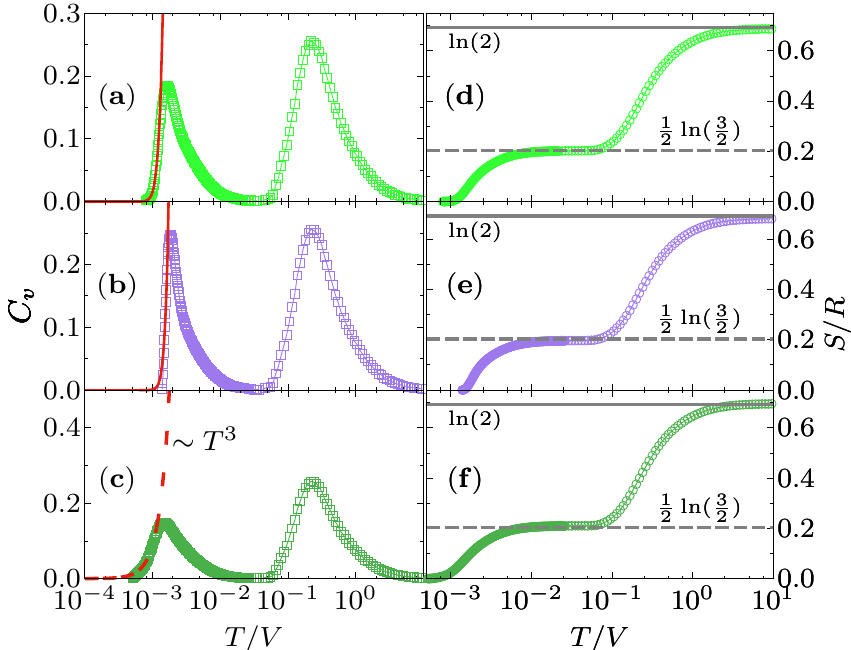}
\caption{{\bf The specific heat and entropy densities for 
different parameter points.} (a) and (d) are the results for 
 point 3 in Figure~\ref{fig1}. (b) and (e) are the results for 
 point 5 in Figure~\ref{fig1}. (c) and (f) are the 
results for point 6 in Figure~\ref{fig1}. On the left panel,
the solid (red) lines indicate the exponential decay and 
the dashed (red) line refers to the power-law behavior.}
\label{supp_Cvs}
\end{figure}

When both ${t_1\neq 0}$ and ${t_2 \neq 0}$, we can have several 
mixed contributions from the $t_1$ and $t_2$ processes. For instance, 
one could apply $t_1$ processes twice and $t_2$ processes twice 
could generate the three-boson collective hopping (see Figure~\ref{hop}). 
All these cases at all orders of perturbation series give positive 
contributions to the collective boson hopping of $t_{\text{coll}}$ and 
thus do not change the sign of $t_{\text{coll}}$ (or the ring exchange 
in the spin language). This justifies the choice of the zero-flux 
sector for the spinon hopping on the diamond lattice in the Method. 
\\

\begin{figure}[t]
\includegraphics[width=0.9\columnwidth]{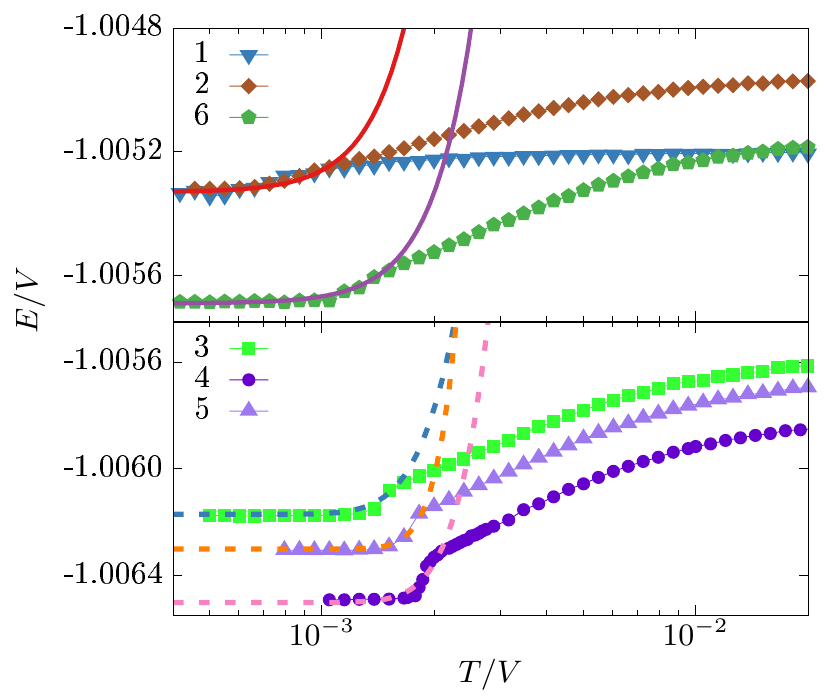}
\caption{{\bf The energy densities for different parameters at low temperatures.} 
The numbers of legends correspond to the points in Figure~\ref{fig1}. 
The solid lines mean ${\sim T^4}$ behaviors and the dash lines mean exponential decay.}
\label{senergy}
\end{figure}

\noindent{\bf More supporting data.} In this part, we list 
additional QMC results to support our conclusion that was made 
in the main text. In Figure~\ref{supp_Cvs}, we plot the specific 
heats $C_v$ and entropy densities $S/R$ of the points 3,5,6 in Figure~\ref{fig1}. 
At low temperatures $C_v$ decays exponentially for the points 3 and 5. 
For the point 6, it is a	power-law decay. There are entropy plateaus 
at the value of Pauling entropy $\frac{1}{2}\ln(\frac{3}{2})$ 
in the plots of the entropy curves. This suggests 
that all these three parameter points experience the degenerate classical 
spin ice manifold during cooling. The energy densities of the parameter points 1-6 
with decreasing temperature are represented in Figure~\ref{senergy}. 
Numerically the energy densities of the parameter points 2,6 show 
power-law behaviors and it is exponentially decaying for the parameter 
points 3,4,5. As for the point 1 the simulation is more difficult so 
the data below $\sim \mathcal{O}(10^{-3} T/V)$ were not so great and we were unable 
to fit it in the plot.
\\
	
\noindent{\bf More discussion about the specific heat.} 
We notice that the low-temperature peak of the specific heat in Figure~\ref{fig4}(c) 
is quite sharp and much sharper than the ones in Figure~\ref{fig4}(b) and/or
Figure~\ref{supp_Cvs}(c). We provide a thermodynamic explanation for this phenomenon. 
We start from the entropy plateau at the value of the Pauling entropy at an 
intermediate temperature, below which the entropy would be gradually lost 
as we cool the system. For a gapped system that is expected for Figure~\ref{fig4}(c),
the entropy loss of the low temperature regime would be relatively small due to the 
energy gap. In contrast, the entropy losses of the low temperature regime for a gapless
case in Figure~\ref{fig4}(b) and Figure~\ref{supp_Cvs}(c) would certainly be more. 
As a result, from the conservation of entropy, the remaining entropy loss would take place
near the low temperature peak, and we must have a larger entropy loss with a higher peak
at the low temperature in Figure~\ref{fig4}(c) to compensate the large remaining entropy.

Unlike the $\mathbb{Z}_2$ liquid in 2D, the $\mathbb{Z}_2$ liquid in 3D supports a 
finite temperature thermal transition from the thermal proliferation of the 
line-like extended excitations. If our proposal of $\mathbb{Z}_2$ liquid does apply
to the narrow region between the Coulomb liquid and the ordered phase, we would 
expect a thermal phase transition. Although we cannot resolve this due to the 
system size and numerical difficulty, it is possible that the low-temperature 
specific heat peak in Figure~\ref{fig4}(c) could be associated with the thermal 
phase transition. 
\\

\noindent{\bf The possibility of charge density wave.} In the narrow region between
the Coulomb liquid and the $\mathbb{Z}_2$ symmetry breaking state, we found a gapped
state. In the main text, we discuss the result from the perspective a gapped
$\mathbb{Z}_2$ liquid state, and indeed our results are consistent with the 
expectation for a $\mathbb{Z}_2$ liquid. Moreover, as we have argued in the 
main text, the spinons of the would-be and/or nearby $U(1)$ Coulomb liquid 
have a very small energy gap, and the strength of the boson pairing 
could simply overcome this gap and gain energies from the spinon pairing. 
Despite that this is a quite reasonable account of the numerical results, 
an alternative explanation may also apply to this gapped regime, and we 
simply describe here. Although we think it is not very likely due to energetic
reason that we explain below, it is possible that, in the narrow region,
the system develops a charge density wave (CDW) order for the hardcore bosons. 
For this to occur, we need to have further neighbor density-density interactions 
to overcome the three-boson collective hopping that is the dominant low-energy
process in the gapped regime. This requires higher order perturbation
than the third order and is suppressed. 
Nevertheless, the CDW, if present, supports gapped excitations and 
a finite temperature thermal transition. The way to distinguish
a $\mathbb{Z}_2$ liquid from a CDW is to measure the density-density 
correlator for a large system to direct detect the translation symmetry 
breaking.

\end{document}